%% file: ms.tex
\documentclass[apj]{emulateapj}

\usepackage{amssymb,amsmath}
\usepackage{ifxetex,ifluatex}
\usepackage{fullpage}
\usepackage{graphicx}
\makeatletter
\ifxetex
  \usepackage[setpagesize=false, 
              unicode=false, 
              xetex,
              colorlinks=true,
              linkcolor=blue]{hyperref}
\else
  \usepackage[unicode=true,
              colorlinks=true,
              linkcolor=blue]{hyperref}
\fi

\newcommand{\um}[0]{$\mu$m}
\newcommand{\hii}[0]{{\sc H\,ii}}
\newcommand{\hi}[0]{{\sc H\,i}}

\begin{document}

\title{The Milky Way Project: Leveraging Citizen Science and Machine Learning to Detect Interstellar Bubbles}
\shorttitle{Brut: Automatic Detection of Interstellar Bubbles}
\slugcomment{Astrophysical Journal Supplement, in press}
\shortauthors{Beaumont et al}
\author{Christopher N. Beaumont$^{1,2}$, Alyssa A. Goodman$^2$, Sarah Kendrew$^3$, Jonathan P. Williams$^1$, Robert Simpson$^3$}
\affil{$^1$Institute for Astronomy, University of Hawai'i, 2680 Woodlawn Drive, Honolulu HI 96822;  beaumont@ifa.hawaii.edu}
\affil{$^2$Harvard-Smithsonian Center for Astrophysics, 60 Garden St., Cambridge MA 02138}
\affil{$^3$Department of Astrophysics, University of Oxford, Denys Wilkinson Building, Keble Road, Oxford OX1 3RH, UK}

\begin{abstract}
We present Brut, an algorithm to identify bubbles in infrared images of the Galactic midplane. Brut is based on the Random Forest algorithm, and uses bubbles identified by $>35,000$ citizen scientists from the Milky Way Project to discover the identifying characteristics of bubbles in images from the \textit{Spitzer Space Telescope}. We demonstrate that Brut's ability to identify bubbles is comparable to expert astronomers. We use Brut to re-assess the bubbles in the Milky Way Project catalog, and find that $10-30\%$ of the objects in this catalog are non-bubble interlopers. Relative to these interlopers, high-reliability bubbles are more confined to the mid plane, and display a stronger excess of Young Stellar Objects along and within bubble rims. Furthermore, Brut is able to discover bubbles missed by previous searches -- particularly bubbles near bright sources which have low contrast relative to their surroundings. Brut demonstrates the synergies that exist between citizen scientists, professional scientists, and machine learning techniques. In cases where ``untrained'' citizens can identify patterns that machines cannot detect without training, machine learning algorithms like Brut can use the \emph{output} of citizen science projects as \emph{input} training sets, offering tremendous opportunities to speed the pace of scientific discovery. A hybrid model of machine learning combined with crowdsourced training data from citizen scientists can not only classify large quantities of data, but also address the weakness of each approach if deployed alone.
\end{abstract}

\section{Introduction}
\label{sec:intro}
Stellar feedback has an important influence on the dynamics and energy balance of the Interstellar Medium (ISM) \citep{Zinnecker07}. Winds and radiation fields from massive young stars can reshape nearby molecular clouds. This interaction can replenish energy dissipated by turbulence, trigger star formation by compressing and collecting gas, and even chemically dissociate or physically disperse molecular clouds \citep{Matzner02}.

Interstellar ``bubbles'' are a primary manifestation of stellar feedback. Young OB stars have sufficient stellar winds and ionizing photon luminosity to sculpt spherical or ring-like cavities in their surrounding molecular clouds. Bubbles in particular are relevant because, compared to collimated outflows, they affect a larger volume of ambient molecular clouds and, compared to supernovae, occur around a larger proportion of stars and persist for a longer period of time \citep{Matzner02, Arce11}.

Unfortunately, due to their complex morphologies, bubbles -- like many features of the interstellar medium -- are difficult to identify and analyze. Existing catalogs of spatially extended bubbles have typically been identified visually \citep{Hu81, Marston96,  Kiss04, Helfand06, Churchwell06, Churchwell07, Konyves07, Simpson12}. This has two main disadvantages. First, it is cumbersome and increasingly infeasible as datasets grow ever larger. Second, manual classification is inherently subjective and non-repeatable; humans are susceptible to fatigue, boredom, and subtle selection biases whose impact on the resulting catalog is difficult to calibrate. The problems associated with manual bubble detection are germane to many analyses with a subjective component. 

Machine learning techniques represent a promising solution to these problems. These techniques aim to construct models that can distinguish between different classes of objects, without domain-specific knowledge of what such objects represent -- in other words, they identify purely statistical differences between different populations of data. While such models are not typically useful as \textit{scientific} models, they can be very effective as \textit{computational} models to perform tasks like classification. 

Our goal in this work is to apply machine learning techniques to the task of bubble detection, and to evaluate the potential of this approach. Using a catalog of known bubbles identified by the citizen scientists of the Milky Way Project \citep{Simpson12}, we ``teach'' an algorithm to identify bubbles in image data from the \textit{Spitzer Space Telescope}. We describe the design of this algorithm, which we call Brut, in Section \ref{sec:method}. In Section \ref{sec:expert}, we use a set of expert classifications to measure Brut's performance at bubble detection. In Section \ref{sec:prob}, we demonstrate that this detector produces useful probabilistic estimates for whether any particular image contains a bubble -- these probabilities correlate well with how astronomers classify similar regions. We use this detector to look for biases and incompleteness in existing bubble catalogs. This analysis yields a new catalog of high-probability bubbles, and we explore how the ensemble properties of this catalog differ from the old catalog. In Section \ref{sec:blind}, we apply Brut to the task of discovering bubbles missing from current catalogs. In Section \ref{sec:next_steps}, we consider how this approach applies more generally to future data analysis problems in astronomy.

\subsection{Previous Work}
\label{sec:previous}

\begin{figure}[h!]
\includegraphics[width=3in]{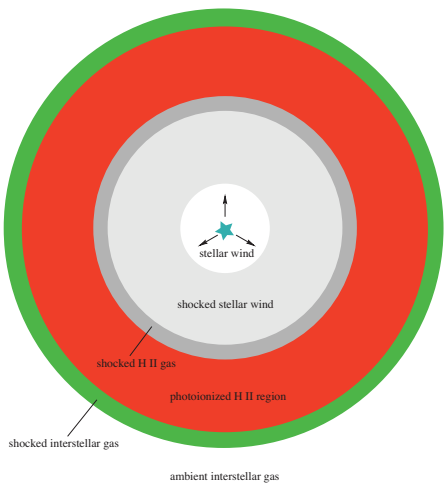}
\caption{Schematic structure of a bubble, adapted from \citep{Freyer03}. The red and green colors encode the typical morphology seen in \emph{Spitzer} images, where green is assigned to 8 \um\, emission -- dominated by PAH fluorescence -- and red is assigned to 24 \um\, emission -- dominated by hot dust.}
\label{fig:schematic}
\end{figure}

Generally speaking, a bubble is a shell-like, 1-30 parsec-scale cavity in the ISM, cleared by a combination of thermal overpressure, radiation pressure, and stellar winds. The basic structure of a bubble is shown in Figure \ref{fig:schematic}. \cite{Stromgren39} first derived the ionization structure around OBA stars in the ISM. Stellar radiation both ionizes and heats gas up to $\sim$ 10$^4$K, creating a strong overpressure that drives expansion into the surrounding medium. This expansion creates an overdense shell of gas along the expansion front. \cite{Castor75} and \cite{Weaver77} extended this analysis to include the effects of strong stellar winds from O and early B stars. The main effect of a stellar wind is to create a shocked, high-temperature (10$^6$K) region within the 10$^4$K ionization region. This shock sweeps up additional material within the ionization region, potentially creating a second shell. Though the term ``bubble'' originally referred to cavities cleared by stellar winds, modern observational studies of ``bubbles'' tend to include both these objects and classical \hii\, regions. Furthermore, \cite{Beaumont10} demonstrated that many bubbles are embedded in relatively thin clouds, and more closely resemble rings than spheres. Throughout this paper, we use the term ``bubble'' broadly to refer to any ring- or shell-like cavity cleared by a young or main sequence star.

The \emph{Spitzer Space Telescope} and its surveys of the Galactic midplane -- {\sc glimpse} \citep{Benjamin03} and {\sc mipsgal} \citep{Carey09} -- enabled comprehensive, statistical studies of bubbles in the Galaxy. Mid-infrared wavelengths are well-suited for bubble observations, as they penetrate deeper into the Galactic disk and match bubble emission features. Figure \ref{fig:schematic} schematically depicts the observational signature of a bubble in \emph{Spitzer} data. The interface between bubble shells and the ambient ISM excites polycyclic aromatic hydrocarbon (PAH) emission features, several of which fall within \emph{Spitzer}'s 8\um\, bandpass. Bubble interiors often emit at 24\um, due to emission from hot dust grains \citep{Everett10}. 

\cite{Churchwell06} and \cite{Churchwell07} carried out the first search for bubbles in \emph{Spitzer} images of the Galactic plane, yielding  a catalog of some 600 bubbles located at $|\ell| < 65^\circ$, $|b| < 1^\circ$. These objects were identified by four astronomers manually searching through 3.5-8.0 \um\, images (they did not have access to 24 \um\, images). \cite{Churchwell06} noted that each astronomer possessed different selection biases, and cautioned that their catalog was likely incomplete.

\begin{figure}[h!]
\includegraphics{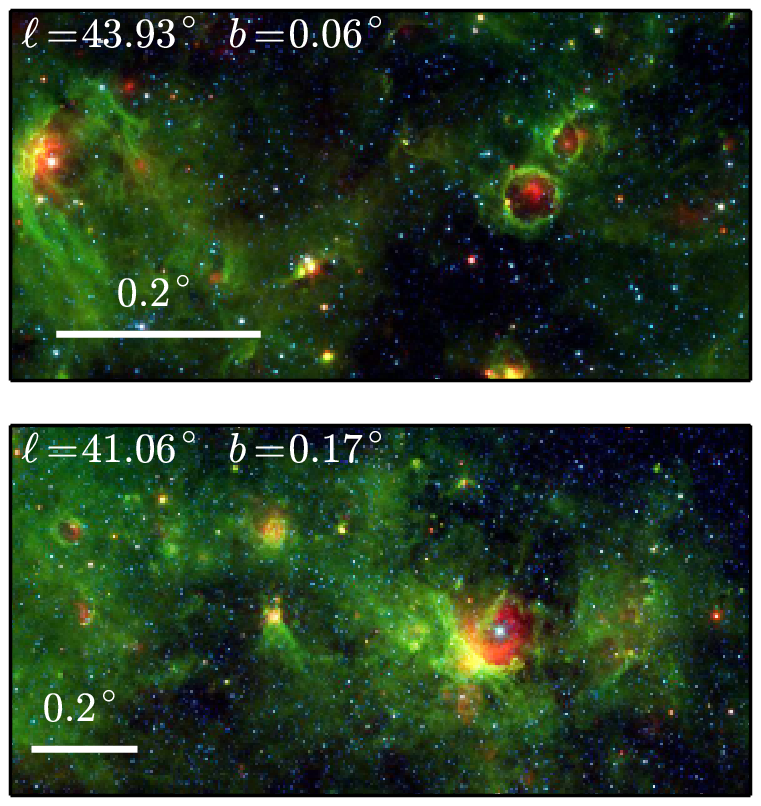}
\caption{Example images presented to Milky Way Project citizen scientists to identify bubbles. The images show 4.5\um\, emission in blue,  8\um\, emission in green, and 24\um\, emission in red.}
\label{fig:mwp}
\end{figure}

In an attempt to overcome the inherent bias in manual classification, the web-based Milky Way Project \citep{Simpson12} enlisted over 35,000 citizen scientists to search for bubbles in \emph{Spitzer} data. The Milky Way Project (hereafter MWP) presented color-composite \emph{Spitzer} images at 4.5\um,  8\um, and 24\um, and asked citizen scientists to draw ellipses around potential bubbles. Figure \ref{fig:mwp} shows two typical images presented to the public. Each image in the MWP was 800x400 pixels, with a pixel scale ranging between 1.35$''$ and 6.75$''$ per pixel. The citizen science effort produced a dramatically larger catalog of $\sim 5000$ objects, nearly 10 times the number in the catalog of $\sim 600$ shells cataloged by the four astronomers of the \cite{Churchwell06, Churchwell07} surveys. The organizers of the MWP attribute this large increase to the 10,000-fold increase in human classifiers and the use of 24\um\, data to further emphasize bubble interiors. They estimate the MWP catalog to be 95\% complete, based on the falloff rate at which new objects were discovered. Still, they cautioned that this catalog is heterogenous, and probably affected by hard-to-measure selection biases. 

\begin{figure*}
\includegraphics[trim= .7in 0 0 0, clip]{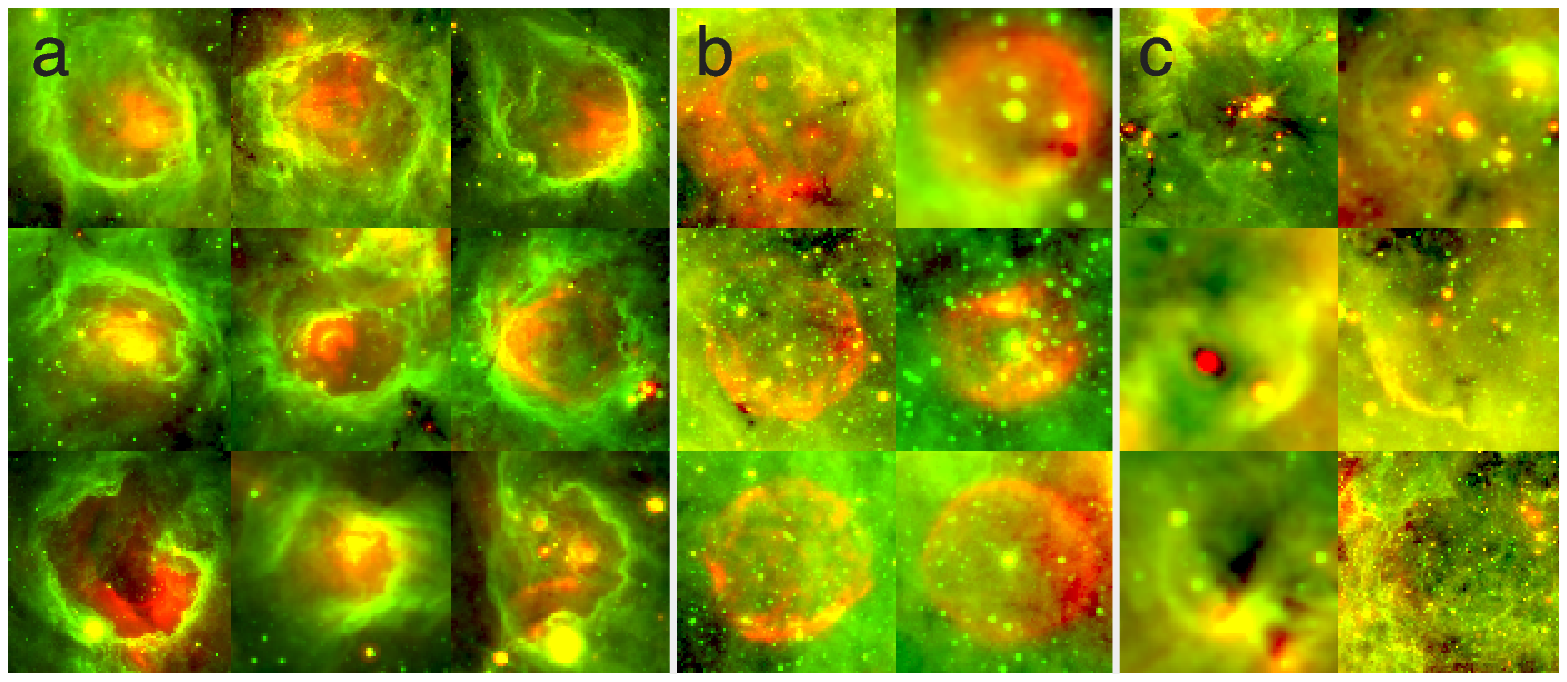}
\caption{Different astrophysical objects in the MWP catalog. a) ``Canonical'' wind-blown bubbles and \hii\, regions. b) shells
without 8 \um\, PAH emission (likely supernovae or bubbles around evolved massive stars). c) generic ISM structures of unclear astrophysical origin.}
\label{fig:gallery}
\end{figure*}

\subsection{Manual and Automatic Classification in Astronomy}
\label{sec:benefits}

In terms of accuracy, humans still outperform computers in most image-based pattern recognition tasks (e.g., \citealt{Zhang10}). Because of this, morphologically complex structures in the ISM (including supernova remnants, outflows, bubbles, \hii\, regions, molecular and infrared dark clouds, and planetary nebulae) are still traditionally cataloged manually. Human classification has several disadvantages, however.

First, human classification is time consuming, and people-hours are a limited resource. Even by enlisting large communities of citizen scientists, data from next generation surveys will be too large to search exhaustively. For example, the $>$ 35,000 citizen scientists of the MWP classified roughly 45 GB of image data from \emph{Spitzer}. Current and next-generation astronomical datasets are many thousands of times larger than this, suggesting tens of millions of citizen scientists would be needed for similar exhaustive searches through tera- and petabyte datasets.

Second, many scientifically important tasks are not suitable for enlisting the public. Part of the appeal of the MWP is due to the fact that the \emph{Spitzer} images are beautiful, contain many bubbles, and are compelling to browse through. Searches for very rare objects, or tasks where the scientific justification is less apparent to a citizen scientist, may be less likely to entice large volunteer communities. \cite{Raddick13} considers the motivations of citizen scientists in greater detail.

Finally, manual classification is not easily repeatable, and hard to calibrate statistically. For example, it is unknown how well the consensus opinion among citizen scientists corresponds to consensus among astronomers. The MWP catalog does not include any estimate of the probability that each object is a real bubble, as opposed to another structure in the ISM. 

Automatic classifications driven by machine learning techniques nicely complement human classification. Such an approach easily scales to large data volumes and is immune to some of the factors that affect humans, like boredom and fatigue. Furthermore, because algorithmic classifications are systematic and repeatable, they are easier to interpret and statistically characterize. Despite the structural complexity of the ISM, \cite{Beaumont11} demonstrated that automatic classification algorithms can discriminate between different ISM structures based upon morphology.

There have been a few previous attempts to detect shell-like structures in the ISM using automated techniques. One approach involves correlating observational data with templates, to look for characteristic morphologies. \cite{Thilker98}, for example, used position-position-velocity templates of expanding shells to search for bubbles in \hi\, data. This technique is also used by \cite{Mashchenko95, Mashchenko02}. \cite{Ehlerova05, Ehlerova13} pursued a more generic approach -- they looked for the cavities interior to \hi\, shells by identifying extended regions of low emission. \cite{Daigle03, Daigle07} searched for \hi\, shells by training a neural network to classify spectra as containing outflow signatures, based on a training set of spectra from known shells. They pass spectral cubes through this classifier, and look for connected regions of outflow signatures. 

It is difficult to robustly classify ISM structures using templates, due to the heterogeneity and irregularity of the ISM -- simple shapes like expanding spheres are often poor approximations to the true ISM. The method pursued by Ehlerov{\'a} et al. is more robust in this respect but, as \cite{Daigle07} noted, is only sensitive to fully-enclosed bubbles. Real bubbles are often broken or arc-like. In this respect the neural network approach that Daigle et al. pursue is an improvement, since neural networks make fewer hard-coded assumptions about the defining characteristics of a bubble, and instead use the data to discover discriminating features. There are a few drawbacks to this approach. First, Daigle et al.'s neural network classifies a single spectrum at a time, with no complementary information about the spatial structure of emission on the sky. Second, the neural networks in \cite{Daigle07} are trained on data from 11 bubbles from the Canadian Galactic Plane Survey, and thus potentially biased towards the peculiarities of this small sample. Finally, neural networks have a large number of tunable parameters related to the topology of the network, and the optimal topology has a reputation for being difficult to optimize. 

Compared to these efforts, the approach we pursue here is similar to the work of \cite{Daigle07}. However, our methodology differs in a few key respects. First, the MWP catalog provides a two orders-of-magnitude larger sample of known bubbles with which to train an automatic classifier. Second, we use the more recent random forest algorithm \citep{Breiman01}, which has a smaller number of tunable parameters that are easier to optimize. Finally, we focus on detecting morphological signatures of shells in \emph{Spitzer} data -- infrared bubbles are visually more prominent than \hi\, shells, and thus potentially easier to classify.

There are several reasons to do build a bubble classifier based on MWP data:

\begin{enumerate}
\item The automatic classifier is capable of producing quantitative reliability estimates for each bubble in the MWP catalog, potentially flagging non-bubble interlopers and leading to a cleaner catalog (Section \ref{sec:prob}).

\item We can search for bubbles not detected by MWP citizen scientists (Section \ref{sec:blind}). 

\item We can treat this task as a case study for complex classification tasks in future datasets, where exhaustive manual classification will not be feasible (Section \ref{sec:next_steps}).
\end{enumerate}

\section{Classification Method}
\label{sec:method}

Our goal is to use the set of citizen-scientist-identified bubbles to build an automatic detector that, when presented with a region of the sky in the  \textit{Spitzer} {\sc glimpse} and {\sc mipsgal} surveys, accurately determines whether or not the image contains a bubble. Our approach, which we name Brut\footnote{\url{http://github.com/ChrisBeaumont/brut}}, is an example of a supervised learning problem. Here is a brief overview of the task:

\begin{enumerate}
\item Build a representative training set of examples of bubble and non-bubble images. This will be derived from the MWP dataset.
\item Convert each example to a numerical \textit{feature vector} that describes each object, and captures the difference between bubbles and non-bubbles.
\item Feed the training data to a learning algorithm to build a model.
\item Use a subset of the examples not used during training to optimize the tunable parameters (so-called \textit{hyper-parameters}) of the learning algorithm.
\end{enumerate}

\subsection{Random Forest Classification}
\begin{figure}[h!]
\includegraphics[width=3in]{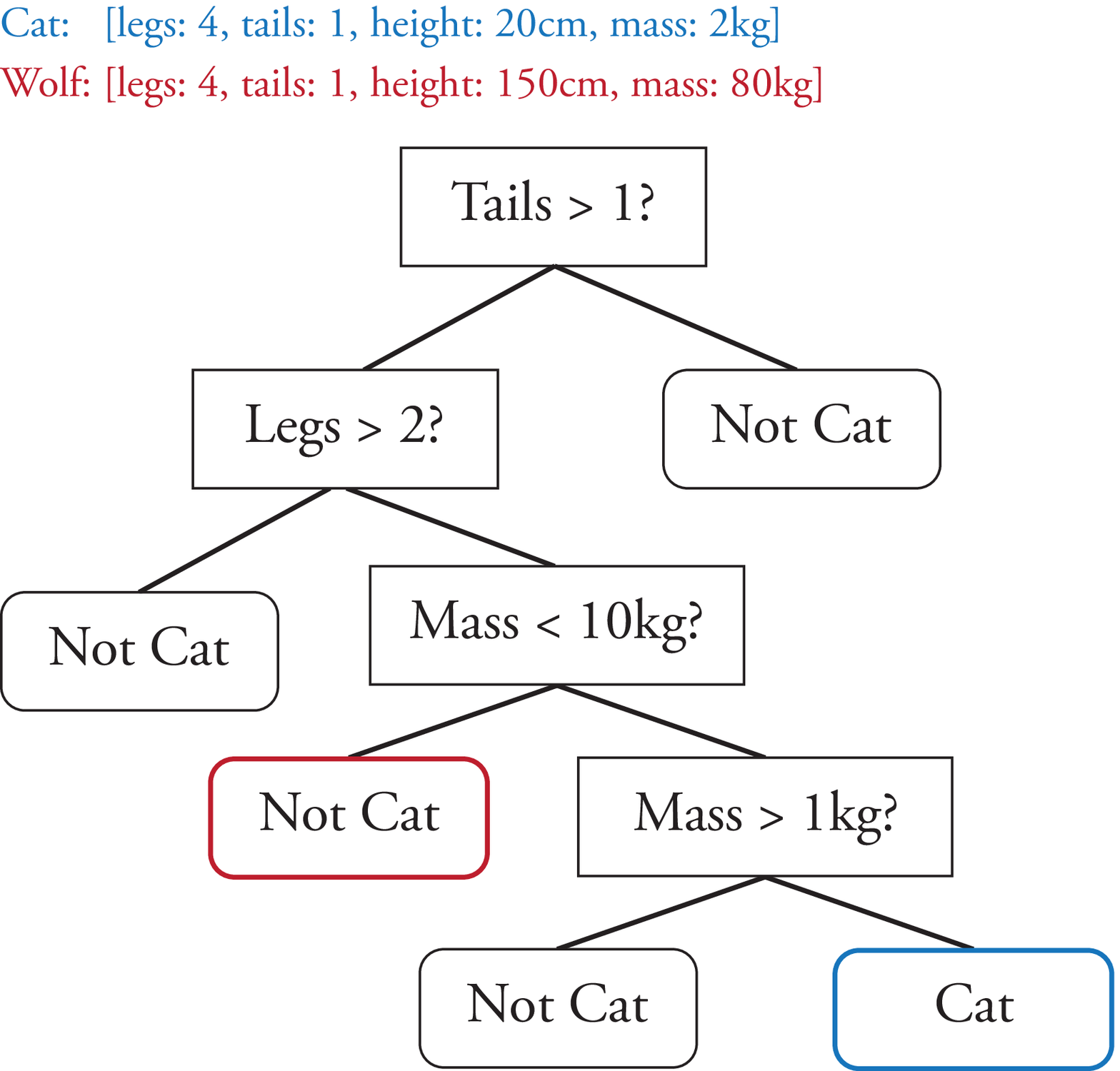}
\caption{Schematic example of a decision tree, to classify observations of animals as cats or not. Two feature vectors are shown at the top, and their classifications are shown as the red and blue outlined nodes of the decision tree.}
\label{fig:decision_schematic}
\end{figure}

Brut uses the Random Forest classification algorithm \citep{Breiman01} to discriminate between images of bubbles and non-bubbles. Random Forests are aggregates of a simpler learning algorithm called a decision tree. A decision tree is a data structure which classifies feature vectors by computing a series of constraints, and propagating vectors down the tree based on whether these constraints are satisfied. For example, Figure \ref{fig:decision_schematic} shows a simple decision tree for classifying animals as cats or not. Here, the feature vectors are four properties of each animal observed: number of legs, tails, height, and mass. Two example feature vectors are shown, and propagate to the outlined classification nodes. In the context of Brut, each classification object is a square region in the {\sc glimpse} survey. Each feature vector is  a set of morphological properties extracted from one such region, and the decision tree predicts whether each region is filled by a bubble.

Decision trees are constructed using an input training set of pre-classified feature vectors. During tree construction, a quality heuristic is used to rate the tree. A few heuristics are common, which consider both the classification accuracy and the complexity of the tree itself -- highly complex trees are more prone to over fitting, and thus disfavored \citep{Ivezic14}. We treat the choice of specific heuristic as a hyper-parameter, which we discuss below. Decision trees are constructed one node at a time, in a ``greedy'' fashion. That is, at each step in the learning process, a new boolean constraint is added to the tree, to maximally increase the score of the quality heuristic. This process repeats until the quality heuristic reaches a local maximum.

On their own, decision trees are prone to over-fitting the training data by adding too many nodes. In a more traditional fitting context, this is analogous to adding too many terms to a polynomial fit of noisy data -- in both situations, the model describes the input data very well, but does not generalize to new predictions. Random Forests were designed to overcome this problem \citep{Breiman01}. Random forests are ensembles of many decision trees, built using different random subsets of the training data. The final classification of a Random Forest is simply the majority classification of the individual trees. Since over-fitting is sensitive to the exact input data used, each individual tree in a Random Forest over-fits in a different way. By averaging the classifications of each tree, the Random Forest mitigates over-fitting. Random Forests have proven effective in many machine learning contexts \citep{Kuhn13}.

In addition to good performance and robustness against over fitting, the Random Forest algorithm has several other advantageous properties: it is possible to interpret how important each element in the feature vector is to the classification (Section \ref{sec:feature_importance}), the algorithm naturally ignores irrelevant elements of the feature vector, and it is conceptually and computationally simple.

It is worthwhile to compare the properties of learning algorithms like the Random Forest classifier to more traditional, ``parametric'' models commonly used in astronomy. Parametric models are often constructed by hand, and derived from an underlying astrophysical model -- the fit to the power spectrum of the Cosmic Microwave Background is a quintessential example of such a model. The structure of a Random Forest model, on the other hand, is set by the clustering properties of the data themselves. The main advantage of the Random Forest approach is that the complexity of the model is to a large extent driven by the complexity of the training data. That is, Random Forests can model complex morphological differences between bubble and non-bubble images, which would otherwise be difficult to express in a manually-constructed model of bubble structure. This is a crucial property when considering the irregularity of the ISM.

\subsection{Feature Extraction}
\label{sec:method_feature_extraction}

The individual ``objects'' that Brut classifies are square regions of the sky, and the goal of the classifier is to determine whether each region is filled by a bubble. Each field is identified by 3 numbers: the latitude and longitude of the center of the field, and the size of the field. We decompose the {\sc glimpse} survey coverage into fields of 18 different sizes, logarithmically spaced from 0.02$^\circ$ to 0.95$^\circ$. At each size scale, we define and classify an overlapping grid of fields. Neighboring tiles in this grid are separated by $1/5$ of the tile size.

\begin{figure}[h!]
\includegraphics{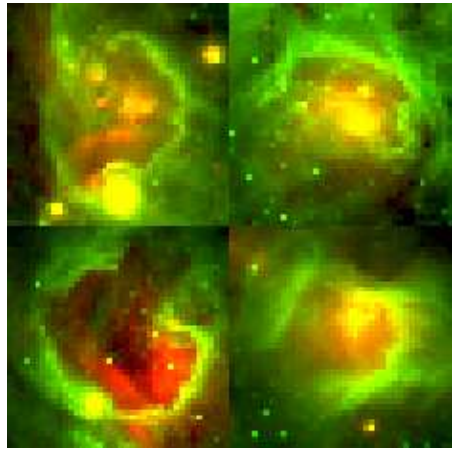}
\caption{Examples of preprocessed images for fields towards four known bubbles.}
\label{fig:preprocess}
\end{figure}

As a preprocessing step, we extracted two-color postage stamps for each field (at 8 \um\, and 24 \um), and resampled these postage stamps to (40x40) pixels. Following a scheme similar to \cite{Simpson12}, these images were intensity clipped at the 1 and 97th percentile, normalized to maximum intensity of 1, and passed through a square root transfer function. The intensity scaling tends to do a good job of emphasizing ISM structure, making bubbles more visible to the eye. Likewise, the (40x40) pixel resolution was chosen because it is reasonably small, yet has enough resolution that postage stamps of known bubbles are still recognizable as such by humans.  Figure \ref{fig:preprocess} shows four preprocessed fields towards known bubbles\footnote{other figures in this paper use the same intensity scaling scheme, but use a (200x200) pixel resolution for presentation purposes}. The goal of preprocessing is to standardize the appearance of bubbles as much as possible, across different size scales and ambient intensities. All subsequent stages of Brut work exclusively with these images, as opposed to the unscaled pixel data.

The input to a Random Forest classifier is a numerical feature vector that describes the properties of each region. The ideal feature vector captures the differences between each class of objects, so that objects from different classes occupy different sub-regions of feature space. Designing an effective feature vector is often the most important determinant of classification performance.

The most obvious choice for a feature vector is simply the numerical value of the pixels in each preprocessed postage stamp. This turns out to be a poor choice, because extended objects like bubbles are characterized by correlations between hundreds or thousands of pixels, and any individual pixel is virtually irrelevant to the classification task. Machine learning algorithms generally perform better when individual elements of the feature vector are more important to the classification, and less dependent on the value of other feature vector elements.

Our feature vectors have been designed based on insights in the automated face detection literature \citep{ViolaJones}.  The basic strategy is to encode a very large number ($\sim$40,000) of generic statistics about image structures at various positions and scales. This strategy seems to be more effective than trying to tune a feature vector to the specific object being identified. While most elements in the feature vector will have no predictive power in the classification, the Random Forest classifier is capable of ignoring this information, and finding the elements in the feature vector with the most predictive power.

The following quantities are extracted from each postage stamp, and concatenated to form a feature vector.

\begin{enumerate}
\item The wavelet coefficients from the Discrete Cosine and Daub4 wavelet transforms \citep{Press07}. These coefficients represent the weights used in representing an image as a linear combination of basis functions. The basis functions in the Discrete Cosine transform are cosine functions, and thus encode information about power at different spatial frequencies. By contrast, the Daub4 wavelets are spatially compact, and thus better encode spatially-isolated structures in an image. These wavelet-derived features resemble the features employed for face detection \citep{ViolaJones}.
\item The image dot product\footnote{i.e., the sum of the pixel-by-pixel product of two images} of each image with 49 template images of circular rings of different size and thickness. Bubbles are morphologically similar to these templates, and tend to have larger dot products than a random intensity distribution. All ring templates are centered within the postage stamp. They span seven radius values ranging from 1-20 pixels, and seven thicknesses (2, 4, 6, 8, 10, 15, and 20 pixels).
\item The byte size of each compressed postage stamp image -- images substantially more or less smooth than a bubble compress to smaller and larger files, respectively.
\item The DAISY features of the image \citep{DAISY}. DAISY features are derived from measurements of local intensity gradients, and encode information about simple shapes in images; they are conceptually similar to edge detection filters.
\end{enumerate}

We arrived at these specific feature vectors through experimentation, because they empirically yield effective classifications. Feature vector design is often the most subjective component of many machine learning tasks, and other features may better model the differences between bubbles and non-bubbles. 

\subsection{Edge-of-field limitations}
\label{sec:feature_edges}
We found that classification performance improved if, in addition to extracting feature vectors at a single $(\ell, b, d)$ field, we also extracted features at $(\ell, b, 2 \times d)$, $(\ell, b, d / 2)$ and $(\ell, b + d/ 2, d)$ -- here, $d$ is the bubble diameter. This ensemble of fields gives the classifier context about the region surrounding a bubble.

Thus, in our approach, Brut extracts feature vector information from a region of the sky twice as large as the field it is classifying. This imposes a latitutde-dependent limit on the largest field Brut is able to classify. Because the {\sc glimpse} coverage is limited to $|b| < 1^\circ$, fields where $|b| + d > 1^\circ$ partially fall outside the survey area. We cannot derive proper feature vectors for these regions, and thus bubbles like this are undetectable by Brut. We return to this fact in Section \ref{sec:prob}.

\subsection{Training Data}
\label{sec:method_training_data}

Building a Random Forest requires providing a set of pre-classified feature vectors. The MWP catalog is a natural choice for building such a training set. However, as shown in Figure \ref{fig:gallery}, the catalog as a whole contains many examples of objects that are not bubbles, and can compromise the training process. Thus, we manually curated a list of 468 objects in the MWP catalog which were clear examples of bubbles. We focus our attention on objects in the large bubble catalog, which contains 3744 out of 5106 total bubbles, with semimajor axes  $a \geq 13.5''$  \citep{Simpson12}. We focus on the large bubble catalog since the web interface used to build the small catalog was different, and may have different selection biases. The largest bubble in the MWP catalog has a semi major axis $a=10.3'$. This upper limit is set by the widest-field images presented to citizen scientists -- we discuss an implication of this cutoff in Section \ref{sec:prob}. The full training set of positive examples is obtained by extracting a field centered on each of these 468 clean bubbles, with a size equal to the major axis of each bubble. 

Training also requires an example set of non-bubbles. In principle, any random region of the sky that doesn't overlap an object in the MWP catalog is a candidate negative example -- while some of these fields actually contain bubbles not in the MWP catalog, they are rare enough that they don't  impact the training. However, we found that selecting negative examples at random was sub-optimal. The reason for this is that most random fields are nearly empty, and easily distinguished from fields with bubbles. Classifiers generated from such a dataset incorrectly labeled nearly any field containing ISM structure as a bubble.

\begin{figure}[h!]
\includegraphics{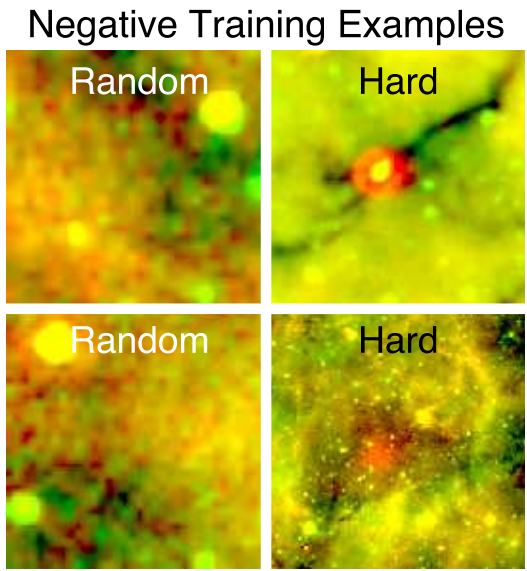}
\caption{Examples of typical randomly selected negative examples (left), and ``hard'' examples (right). Including more hard examples produces more discriminating classifiers.}
\label{fig:bootstrap_neg}
\end{figure}

A better set of negative examples includes more fields containing structure (Figure \ref{fig:bootstrap_neg}). We built such a collection in a bootstrapped fashion. We began with a random set of negative fields, distributed uniformly in latitude and longitude, with sizes drawn from the size distribution of the positive training set. We trained a classifier with these examples and used it to scan 20,000 bubble-free regions. We then discarded half of the initial negative examples (those classified most confidently as not containing bubbles), and replaced them with a random sample of the mis-classified examples. We repeated this process several times, but found that one iteration was usually sufficient to build a good set of training data.

Our final training set consists of 468 examples of bubbles identified by the Milky Way Project, and 2289 examples of non-bubble fields. We found that an approximate five-fold excess of negative examples yielded a more discriminating classifier with lower false positive rates, compared to a training set with equal number of positive and negative examples. Likewise, using dramatically more negative examples compromised the classifier's ability to positively identify known bubbles. The number of positive examples was set by the number of clear bubbles in the Milky Way Project catalog itself (as judged by the lead author).

\subsection{Partitioning the Classifier}
\label{sec:partition}
Instead of building a single Random Forest classifier, we trained three forests on different subsets of the sky. Each forest was trained using examples from 2/3 of the {\sc glimpse} survey area, and used to classify the remaining 1/3. The motivation for doing this is to minimize the chance of over-fitting, by ensuring that the regions of the sky used to train each classifier do not overlap the regions of sky used for final classification. 

\begin{figure}[h!]
\includegraphics{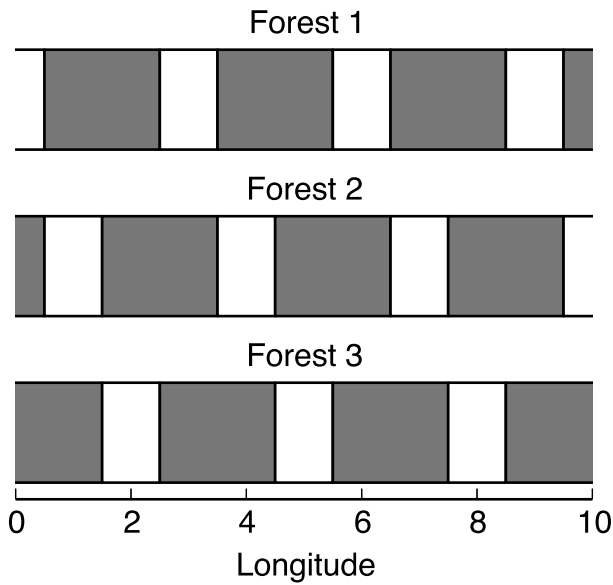}
\caption{Illustration of the zones used to train each Random Forest. Each forest is trained using the data from the shaded regions in its zone, and used to classify the light regions.}
\label{fig:zone}
\end{figure}

The training zones used to classify each forest are interleaved across all longitudes. For example, Figure \ref{fig:zone} depicts how the $0^\circ < \ell < 10^\circ$ region is partitioned -- the shaded regions denote the portion of the sky used to train each forest, and the white region shows the zone each forest is responsible for classifying.

\subsection{Hyperparameter Optimization}
The Random Forest algorithm has a small number of tunable parameters:

\begin{enumerate}
\item  The heuristic used to measure classification quality when building a decision tree. We considered three heuristics: the Gini impurity, information gain, and gain ratio criteria. All of these provide similar measurements of how well a Decision tree partitions a training dataset -- \citep{Ivezic14} discuss these criteria in greater detail.
\item The number of individual decision trees to include in the Random Forest. We considered forests with 200, 400, 800, and 1600 trees.
\item A stopping criterion, when building individual decision trees. If this criteria is set to one, new nodes will always be added to a tree, until the quality heuristic reaches a local maximum. If this number is set to a value $c > 1$, then it prevents a subtree from being added to a given node in the decision tree unless at least $c$ training examples propagate to that node. This criteria can suppress over-fitting in individual trees, though it seems to be less important in the context of Random Forests. We considered values of 1, 2, and 4.
\end{enumerate}

We explored the impact of these hyperparameter settings via cross validation. During cross validation, we split the training examples into two groups: a primary set with 154 bubble examples and 1462 non-bubble examples, and a validation set of 157 bubble and 10000 non-bubble examples. We trained a Random Forest with a particular choice of hyper parameters, and measured the accuracy and false-positive rate of the classifier on the validation set. We found that a forest using the Information Gain heuristic, 800 trees, and $c=4$ yielded the best performance. The quality heuristic and forest size are stronger determinants of classification performance than $c$. Classifier performance largely converged at 800 trees. While forests with 1600 trees did no worse than 800-tree forests, classification time was twice as slow. Once we converged on an optimal set of hyperparameters, we trained the final three Random Forest classifiers using the scheme discussed above.

\subsection{Feature Importance}
\label{sec:feature_importance}
The Random Forest algorithm measures the importance of each element in the feature vector during training. Intuitively, elements in the feature vector which appear in a large number of nodes (see Figure \ref{fig:decision_schematic}) are more important to the classification. Likewise, elements which appear closer to the root of the tree are more important, since a larger fraction of examples propagate through those nodes. More formally, feature importance is defined by the average improvement in the quality heuristic for all nodes associated with a feature, weighted by the number of training examples which propagate through those nodes. Feature importance is more meaningful in a relative sense than an absolute sense -- features with higher importance scores have a greater impact on the behavior of the decision tree.

Table \ref{tab:importance} summarizes the importance of each category of features used in Brut -- it shows the average and sum of the importances for all feature elements in each category. Both columns are normalized by the maximum importance value. The features derived from ring templates have the highest total importance -- this is further evidenced by the fact that 19 out of the 20 most important single features are ring features (the 15th most important feature is a wavelet feature). While the average importance of the compression feature is slightly greater, there are far fewer compression features than ring features. 

\begin{deluxetable}{lrr}
\tablecolumns{3}
\tablewidth{0in}
\tabletypesize{\scriptsize}
\tablecaption{Feature importances}
\tablehead{\colhead{Feature Category} & \colhead{Mean Importance} & \colhead{Total Importance}}
\startdata
ring	&  0.945 & 	 1.000 \\
compression &  1.000	 &  0.014 \\
wavelet &  0.010 	&  0.292\\
daisy &	 0.007	&  0.085 \\
\enddata
\label{tab:importance}
\end{deluxetable}

\subsection{Using the classifier}
\label{sec:using_classifier}

To classify a region after training, we compute the feature vector and dispatch it to one of the three random forests (depending on longitude, as discussed in Section \ref{sec:partition}). The forest produces a classification score between -1 and 1. This number is equal to the fraction of trees in the forest which predict the feature vector is a bubble, minus the fraction of trees which predict it is not. This score provides more information than a simple binary classification, as it gives some sense of the confidence of the classification. Furthermore, one can adjust the threshold that defines when an object is considered to be a bubble. When building a set of bubbles detected by Brut, increasing this threshold removes bubbles with lower classification scores. Ideally this increases the reliability of the catalog -- since the remaining bubbles have higher scores and judged by the classifier to be clearer bubbles -- at the cost of completeness. 

\begin{figure}[h!]
\includegraphics{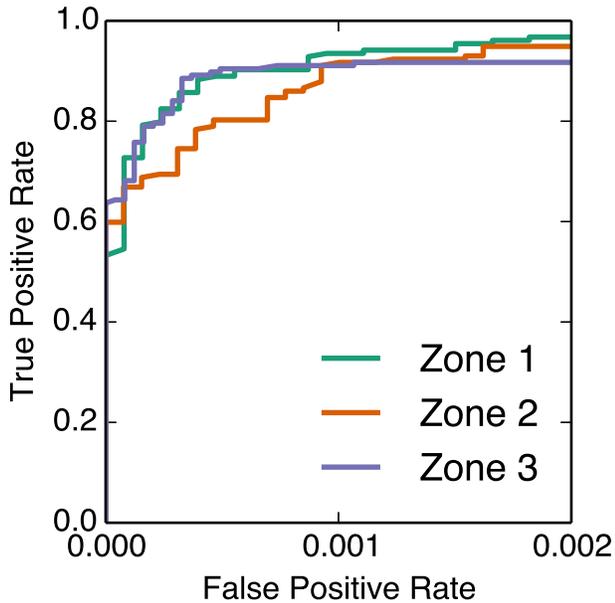}
\caption{The tradeoff between false positive rate and true positive rate, for the three Random Forests in Brut. Each classifier is trained using a different zone, depicted in Figure \ref{fig:zone}.}
\label{fig:roc}
\end{figure}

One standard way to summarize the performance of a classifier is to plot the false positive rate (fraction of negative examples incorrectly classified) versus the true positive rate (fraction of bubbles correctly classified) as the threshold varies. This is called the Receiver Operating Characteristic, and is shown for the three classifiers in Figure \ref{fig:roc}. The false positive rate is measured by classifying $\sim 50,000$ random negative fields, and the true positive rate is measured using the 468 known bubbles in the training set -- note however that, given our partitioning strategy, each of these bubbles is classified by a forest that has \textit{not} seen this example before. Also recall that the set of bubbles in this test set were selected because they are very well defined. Thus, Figure \ref{fig:roc} overestimates the true positive rate when presented with more ambiguous bubbles. We compare the performance of the classifier on more representative bubbles in Section \ref{sec:expert}.

The second forest has a slightly worse performance curve -- at a true positive rate of 80\%, the forest that classifies zone 2 has a false positive rate about twice that of the other forests. In absolute terms, this false positive rate of 0.0005 corresponds to six false positives out of the 13,000 test examples classified by this forest. 

\subsection{Building a Catalog}
\label{sec:build}

Using our automatic bubble classifier to conduct a full search for bubbles in \textit{Spitzer} data is straightforward. The main task involves scanning the classifier across the {\sc glimpse} survey footprint at different size scales, as described in Section \ref{sec:method_feature_extraction}. The classifier assigns a score to each field of view, as discussed in the previous Section. To convert this continuous score into a binary ``bubble/not-bubble'' classification, we can choose a numerical threshold, and consider any score above this threshold to be a bubble detection. From Figure \ref{fig:roc}, the choice of this threshold sets the tradeoff between completeness (high true positive rate) and reliability (low false positive rate). Because different scientific applications
have different requirements on completeness and reliability, it's useful to set this threshold on a per-analysis basis. As we also discuss shortly, we can turn these confidence scores into probabilistic estimates that an expert would consider the region to be a bubble, by calibrating against expert classifications. 

Note that Brut only attempts to determine the location and size of bubbles. Citizen scientists, on the other hand, were able to constrain the aspect ratio, position angle, and thickness of each object. 

\begin{figure}[h!]
\includegraphics{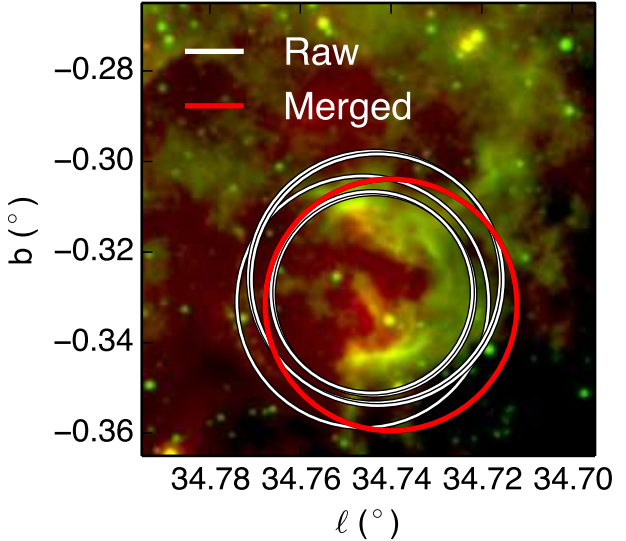}
\caption{An illustration of how Brut merges multiple bubble detections. Each white circle is a location with a high Brut score. The red circle is the region with the highest brut score, and is used as the final detection.}
\label{fig:cluster}
\end{figure}

In practice, the classifier is insensitive to small adjustments to the size or location of the field. As a result, bubbles are usually associated with a cluster of fields that rise above the detection threshold (the white circles in Figure \ref{fig:cluster}). We follow a simple procedure to merge these individual detections. Given a collection of raw detections, this procedure identifies the two nearest locations at similar scales, and discards the location assigned a lower score by the classifier. This process repeats until no two fields are close enough to discard. Two fields are close enough to merge if they are separated by less than the radius of either field, and have sizes that differ by less than 50\%. The red circle in Figure \ref{fig:cluster} shows the result of merging detections.

This practice works well for isolated bubbles, but occasionally fails for bubble complexes with multiple, overlapping bubbles. A more sophisticated clustering analysis may well perform better in this situation.

\section{Expert Validation}
\label{sec:expert}

To evaluate the performance of Brut, we need a set of ``ground truth'' classifications. However, to some extent identifying bubbles in \textit{Spitzer} images is inherently subjective. While the objects in Figure \ref{fig:gallery}a are unambiguously associated with star formation, other sources are not. In some cases -- particularly irregular bubbles, or those with faint or no 24 \um\, emission -- it is unclear whether cavities seen in 8 \um\, data are actively sculpted by young stars, or merely coincidentally ring-like structures in the ISM.

The subjectivity of bubble identification has received only limited attention. The MWP catalog includes a ``hit rate'' measurement that lists the fraction of citizen scientists who identify a specific bubble, relative to the total number of users who viewed an image containing that bubble. However, as discussed above, the MWP catalog contains many objects besides bubbles. Thus, while the hit rate communicates how visually salient a particular ISM feature appears to MWP users, it does not directly convey how much consensus astronomers have about the astrophysical nature of that feature.

To better measure the level of expert consensus in bubble identification, we conducted a small online survey. The astronomer participants of this survey were presented with a sequence of 92 \emph{Spitzer} images at three zoom levels and two contrast settings. They were asked to assign each image to one of three categories: clear examples of bubbles or \hii\, regions, ambiguous or irregular bubbles, or Non-bubbles. Appendix A discusses the survey setup in more detail. We chose approximately 100 test images to keep the exercise short, and minimize fatigue. We further minimized the effects of fatigue by presenting examples to each expert in random order.

\subsection{Validation of the Milky Way Project Catalog}
\label{sec:expert_mwp}
Of the 92 images in the expert survey, 45 were a random subset of objects in the MWP catalog (the remaining fields are discussed in the next section). Figure \ref{fig:expert_mwp_votes} shows the voting breakdown for these objects. Each row corresponds to a single image, and the red, orange, and blue bars show the fraction of experts who classified the image in each category.  We call attention to several aspects of this distribution:

\begin{figure}[h!]
\includegraphics{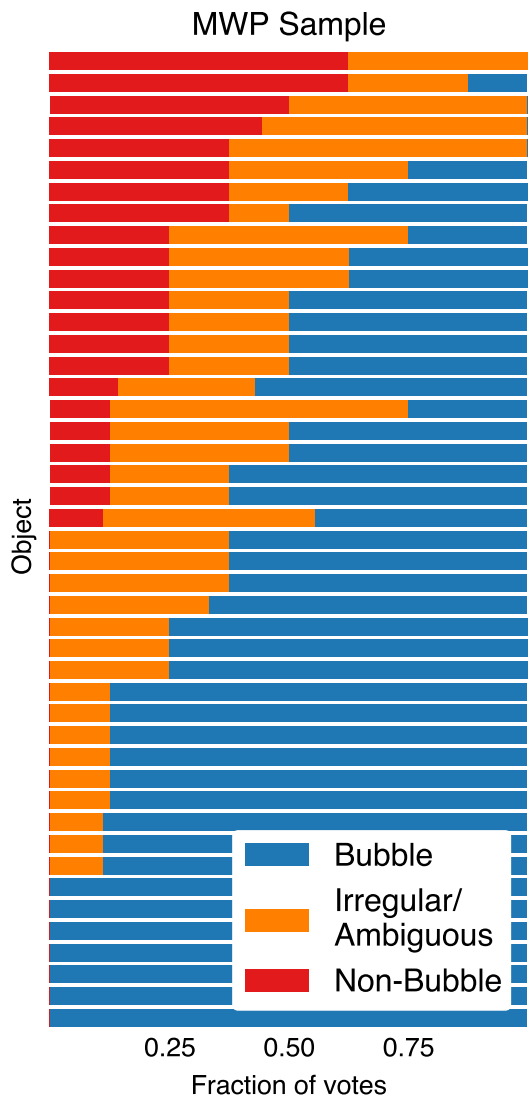}
\caption{The distribution of expert classifications for 41 objects in the MWP bubble catalog. Each row corresponds to an image, and the colored bars in that row indicate the fraction of experts who categorized the image as a Bubble (blue), Irregular or Ambiguous Bubble (orange), and Non-Bubble (red).}
\label{fig:expert_mwp_votes}
\end{figure}

\begin{enumerate} 
\item Only 15\% of these objects -- 7 out of 45 -- were unanimously classified as clear bubbles -- this number increases to about 50\% if all irregular and ambiguous objects are optimistically treated as bubbles. 
\item About 25\% of the objects -- 12 out of 45 -- are judged by a plurality of experts to be irregular, ambiguous, or non-bubbles -- that is, objects which are less likely to be the result of a young star sculpting the surrounding ISM. While some of these regions might be other interesting astrophysical objects like supernova remnants, most appear to be coincidental arc-like structure in the ISM, not convincingly sculpted by young stars.
\item The most popular category for each object is chosen by 70\% of experts on average. In other words, when using \emph{Spitzer} images, determining the ``true'' nature of a typical object in the MWP catalog is subjective at the 30\% level. Citizen scientists, by comparison, exhibit much lower levels of agreement; the average hit-rate for these objects is 20\%.
\end{enumerate}

\begin{figure*}
\includegraphics{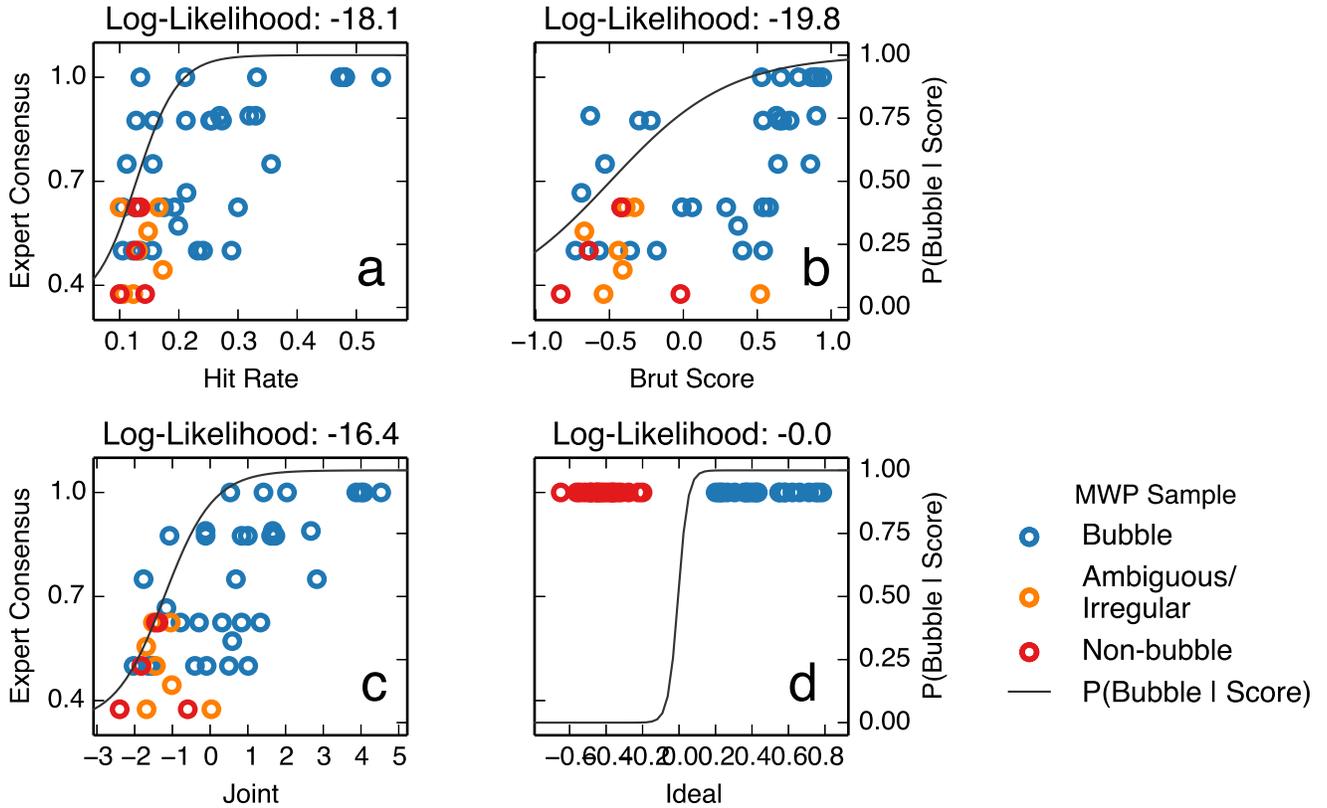}
\caption{The ability for the MWP hit rate and Brut score to predict expert classifications for objects in the MWP catalog. Each circle represents one of the 41 objects in the MWP catalog shown to experts. Color depicts the most popular classification, and the y-position indicates the fraction of experts who chose this category. The curves estimate the bubble probability (right Y axis), obtained via logistic regression to the object category. The Joint metric, which combines the Hit Rate and Brut score, is best able to predict expert classifications. The title reports the log-likelihood of each logistic fit, according to Equation \ref{eq:likelihood}. Panel d depicts the ideal situation, where classes are unambiguous.}
\label{fig:expert_mwp_score}
\end{figure*}

Is it possible to determine which of the remaining objects in the MWP catalog would likely be rejected as interlopers by experts? The hit rate of each object in the catalog might correlate with expert opinion, and be a useful filter. Likewise, the confidence score computed by Brut might identify interlopers. Figure \ref{fig:expert_mwp_score} plots the citizen's hit rate and Brut's confidence score compared to the expert classification. The Y position of each point denotes the level of consensus for each object -- the percentage of experts who chose the plurality category. Both metrics partially separate bubbles from the other categories. The hit rate is more effective at penalizing orange and red objects, which are confined to hit rate $<0.2$ in panel a. On the other hand, the hit rate is ineffective at isolating bubbles with high expert consensus. The Brut score is more effective at identifying these high-consensus bubbles, which is apparent by the cluster of blue points in the upper right corner of panel b. We also plot in Figure \ref{fig:expert_mwp_score}c a joint score -- the sum of the normalized and mean-subtracted Brut score and hit rate.  This combines the strengths of the individual metrics, and achieves the best separation.

The expert reclassifications of this sample of MWP objects can be used to convert a raw score like the joint score to a calibrated probability that an expert would classify an object as a bubble, given that score. To achieve this, we perform a logistic regression against each of the three scores (hit rate, Brut score, and joint score). The logistic regression only considers whether the plurality category for each object is a bubble -- the consensus information is not used. The grey curves in Figure \ref{fig:expert_mwp_score} trace this curve, and show the predicted probability that an object is a bubble given the score on the X axis. We reiterate that this is \emph{not} a fit through the (X, Y) cloud of points in the plot; rather, it is (informally) a fit to the fraction of circles which are blue at each X location of the plot. The title of each panel reports the quality of each fit, as defined by the log-likelihood:

\begin{align}
\mathcal{L} & = \sum_{\rm Bubbles} \log\left(P\left({\rm Bubble | Score}\right)\right) + \nonumber \\ 
 & \sum_{\rm Non-Bubbles} \log\left(1 - P\left({\rm Bubble | Score}\right)\right) 
\label{eq:likelihood}
\end{align}

Panel d in Figure \ref{fig:expert_mwp_score} shows the ideal situation, where a scoring metric perfectly separates two classes of unanimously-categorized objects. The joint score combines the information from citizen scientists and Brut, and comes closest to this ideal. While the logistic curves for the hit rate and joint score look similar, the latter curve is a better fit to the data as judged by the higher likelihood of the data under this model.

\subsection{Uniform Classification}
The remaining 47 regions in the expert survey were selected randomly from the full set of fields analyzed during Brut's full scan of the \emph{Spitzer} images. A fully random sample from this collection is uninteresting, since the majority of fields are blank areas of the sky. Instead, these 47 images are approximately uniformly distributed in the Brut confidence score. We refer to these images as the ``uniform'' sample. 

The expert vote distribution for these images is shown in Figure \ref{fig:expert_uniform_votes}. There are more non-bubble fields in this sample, but the overall properties are otherwise similar. The average field is classified with a 72\% consensus.

Figure \ref{fig:expert_uniform_score} shows the Brut score for each field in the uniform sample, compared to the plurality category (color) and level of consensus (Y axis). The Brut score does a good job of separating the high consensus objects. Overall, however, the classes overlap more for the uniform sample than for the MWP sample. This is also expected, since all objects in the MWP sample have been previously identified by citizen scientists as potential bubble sites. The uniform sample does not benefit from this information, and is representative of the harder, ``blind'' classification task. As before, a logistic fit can be used to estimate, for objects identified in a blind search, the probability of being a bubble given the Brut score.

\begin{figure}[h!]
\label{fig:calibration}
\includegraphics[width=3in]{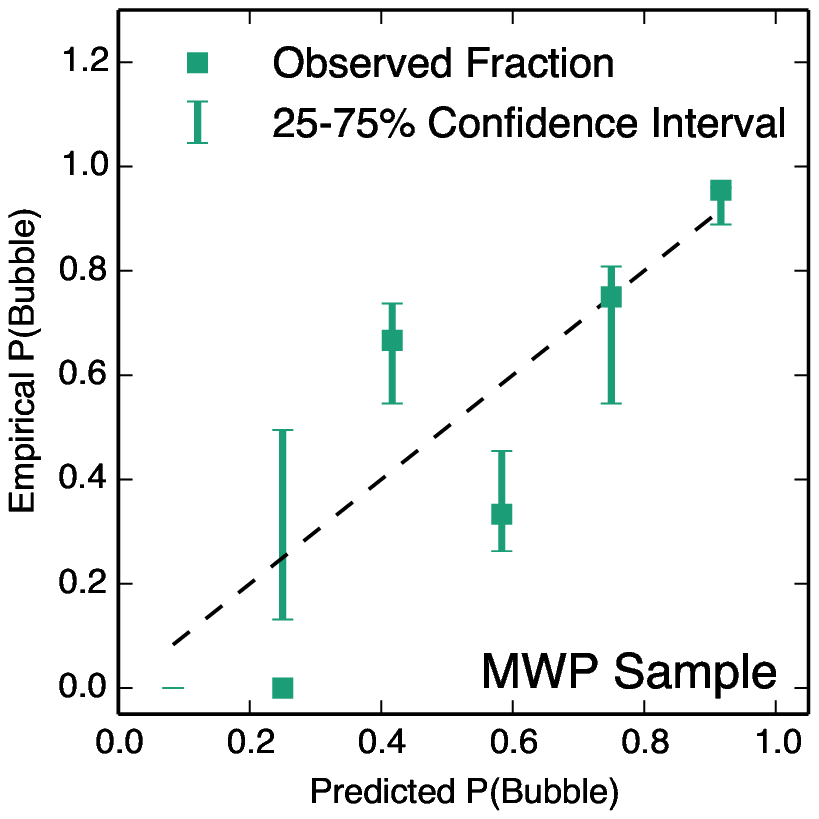}
\includegraphics[width=3in]{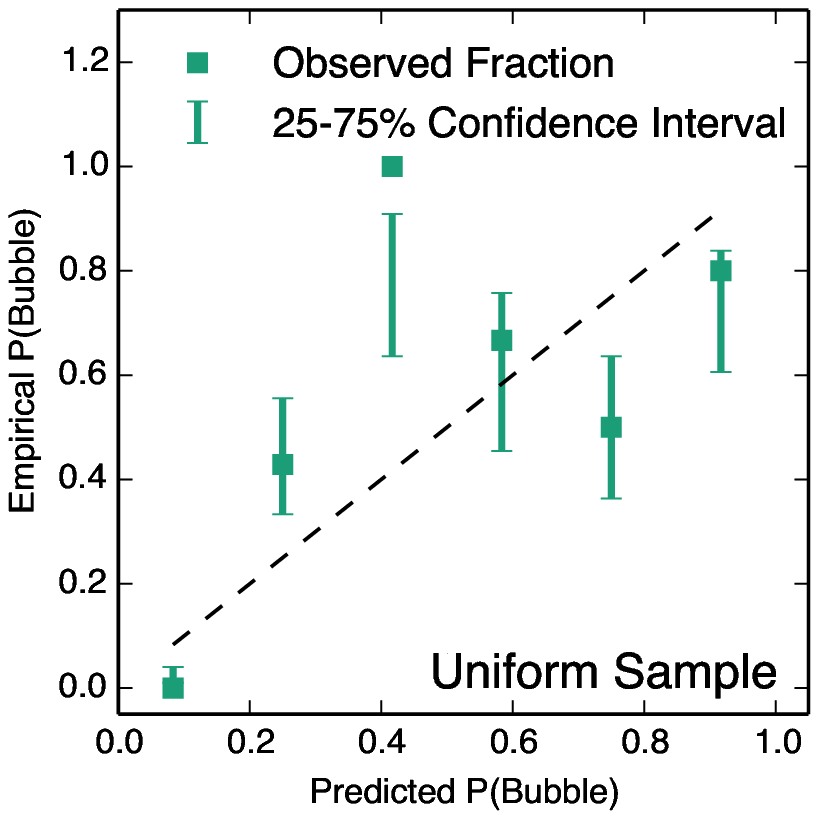}
\caption{Calibration plots for the logistic fit to the expert votes in the MWP sample (left) and Uniform sample (right). Each point represents a collection of samples binned by predicted bubble probability (X axis). The empirical bubble fraction is shown by the
datapoint, and the error bars give the 25-75\% posterior probability interval given the finite sample size.}
\end{figure}

\begin{figure}[h!]
\includegraphics{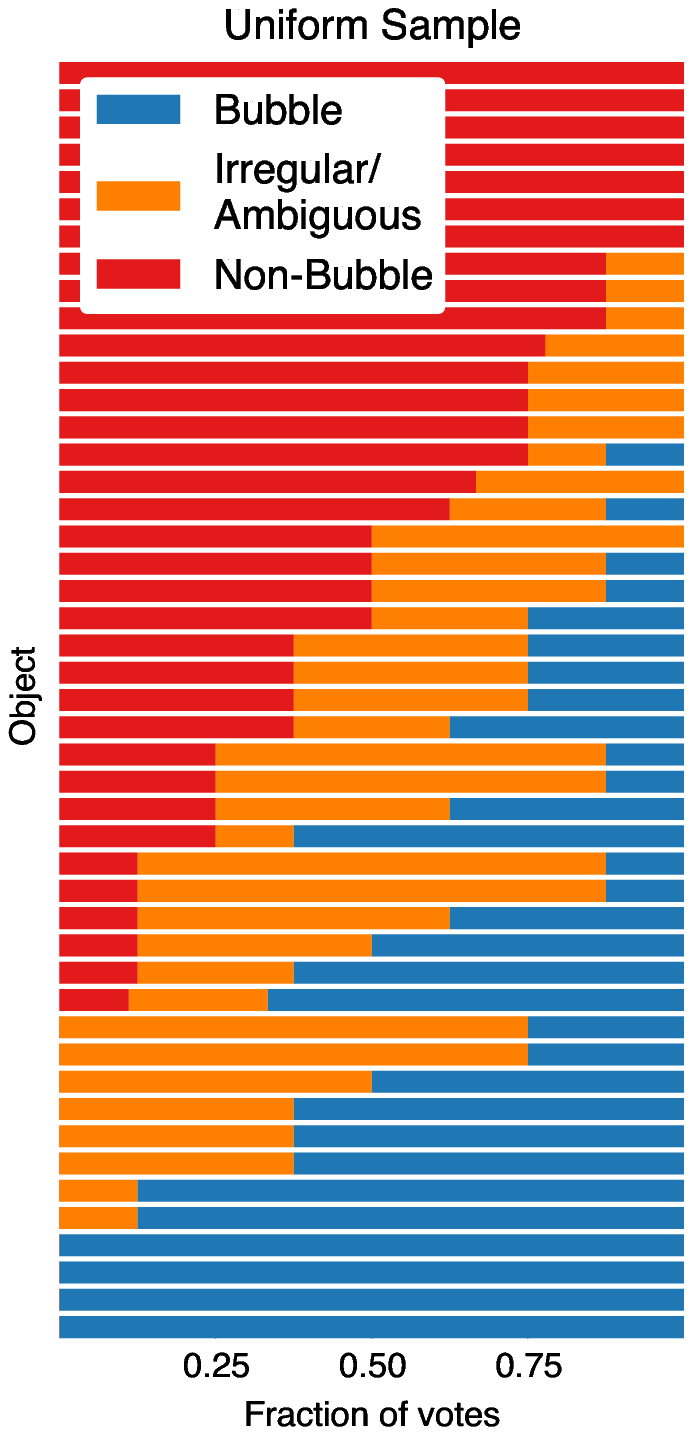}
\caption{The same as Figure \ref{fig:expert_mwp_votes}, for the Uniform sample.}
\label{fig:expert_uniform_votes}
\end{figure}

\begin{figure*}
\includegraphics{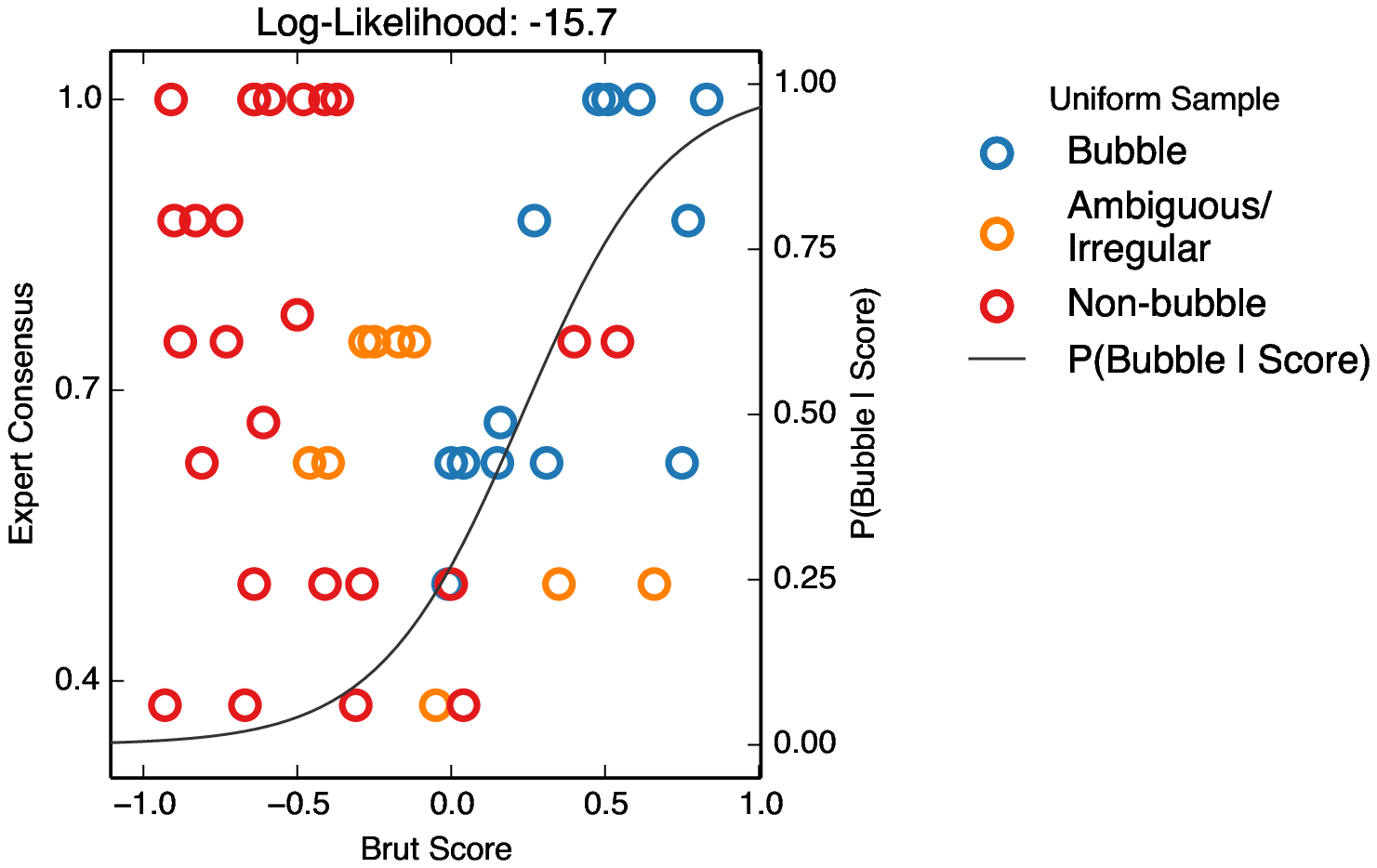}
\caption{The same as Figure \ref{fig:expert_uniform_score}, for the Uniform sample.}
\label{fig:expert_uniform_score}
\end{figure*}

We intend to use the logistic fits in Figure \ref{fig:expert_mwp_score} and \ref{fig:expert_uniform_score} to convert Brut's score into a probabilistic estimate that an expert would judge an image to contain a bubble. A necessary condition for this is that the model be calibrated to the data -- that $P(Bubble | Score)$  is consistent with the empirical bubble fraction as a function of score, for the objects in the expert survey. Figure \ref{fig:calibration} confirms this to be the case -- it divides the MWP and uniform samples into bins of probability (given by the logistic model), and shows the empirical bubble fraction for each bin on the Y axis. Here, bubbles are defined to be objects voted as bubbles by a plurality of experts. The 25-75\% posterior probability band is also shown, derived from the finite sample size assuming a binomial distribution (note that the empirical averages can fall outside this interval). The dotted one-to-one line is the expected performance of a calibrated model, with which these models are broadly consistent.

\section{A Probabilistic Bubble Catalog}
\label{sec:prob}
Section \ref{sec:expert_mwp} outlines a strategy to combine hit rates with Brut's own confidence score to predict whether or not
an expert would classify each object in the original MWP catalog as a bubble. We have assigned a bubble probability to each
object in MWP catalog based on this analysis (Table \ref{table:prob_table}). The high-probability subsample of this catalog displays several interesting differences compared to the unfiltered catalog. 

\begin{figure*}[h!]
\includegraphics{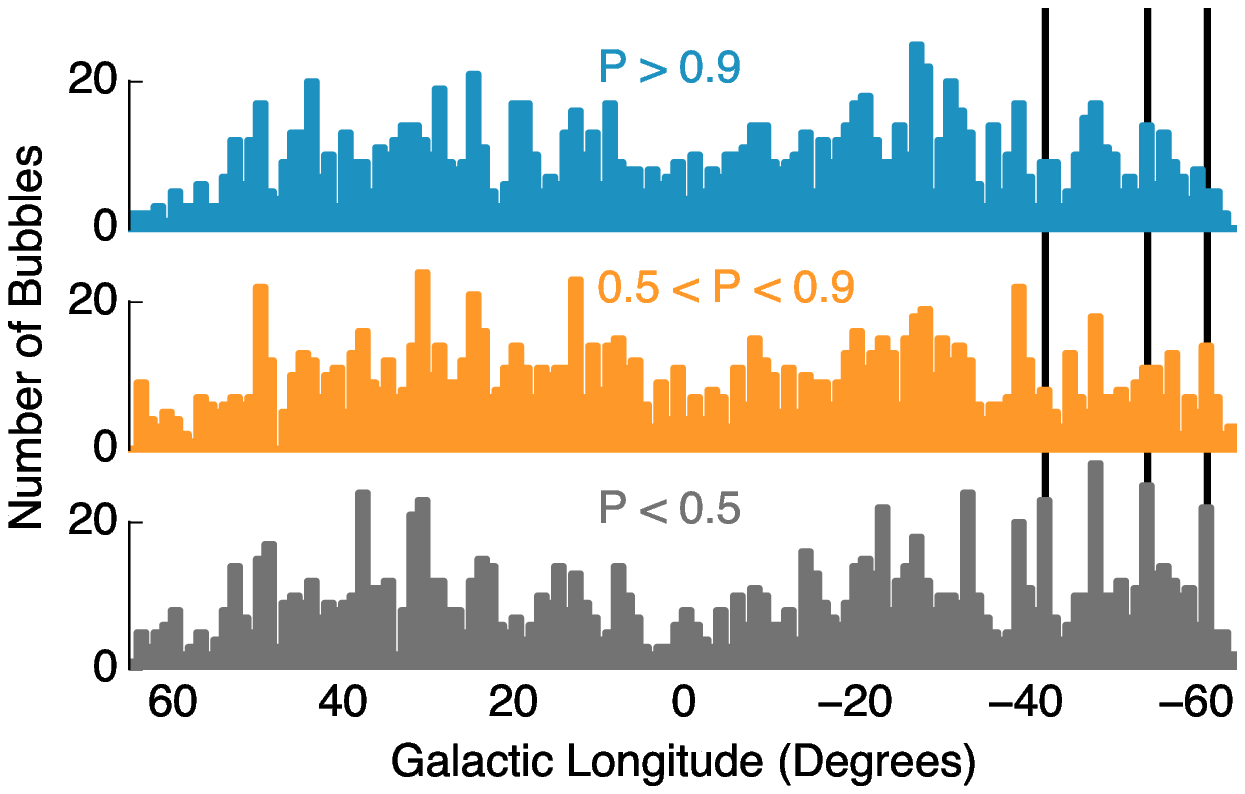}
\caption{The longitude distribution for the MWP catalog, partitioned according to the bubble probability score.
Vertical black lines identify three degree-scale emission complexes with an overabundance of low-probability objects.}
\label{fig:dist_lon}
\end{figure*}

\begin{figure*}
\includegraphics[trim=0 .3in 0 .6in, clip]{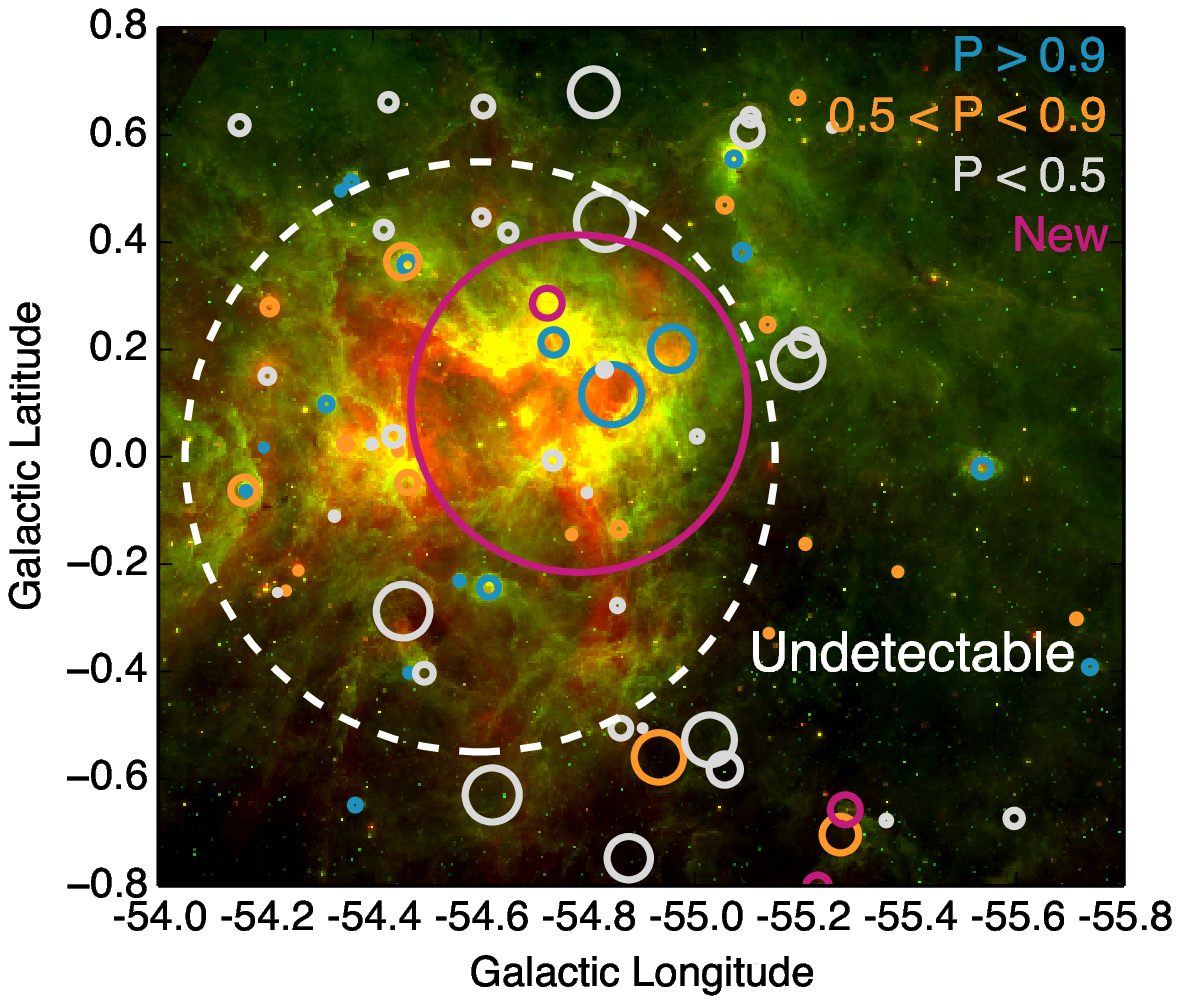}
\includegraphics[trim=0 .3in 0 .6in, clip]{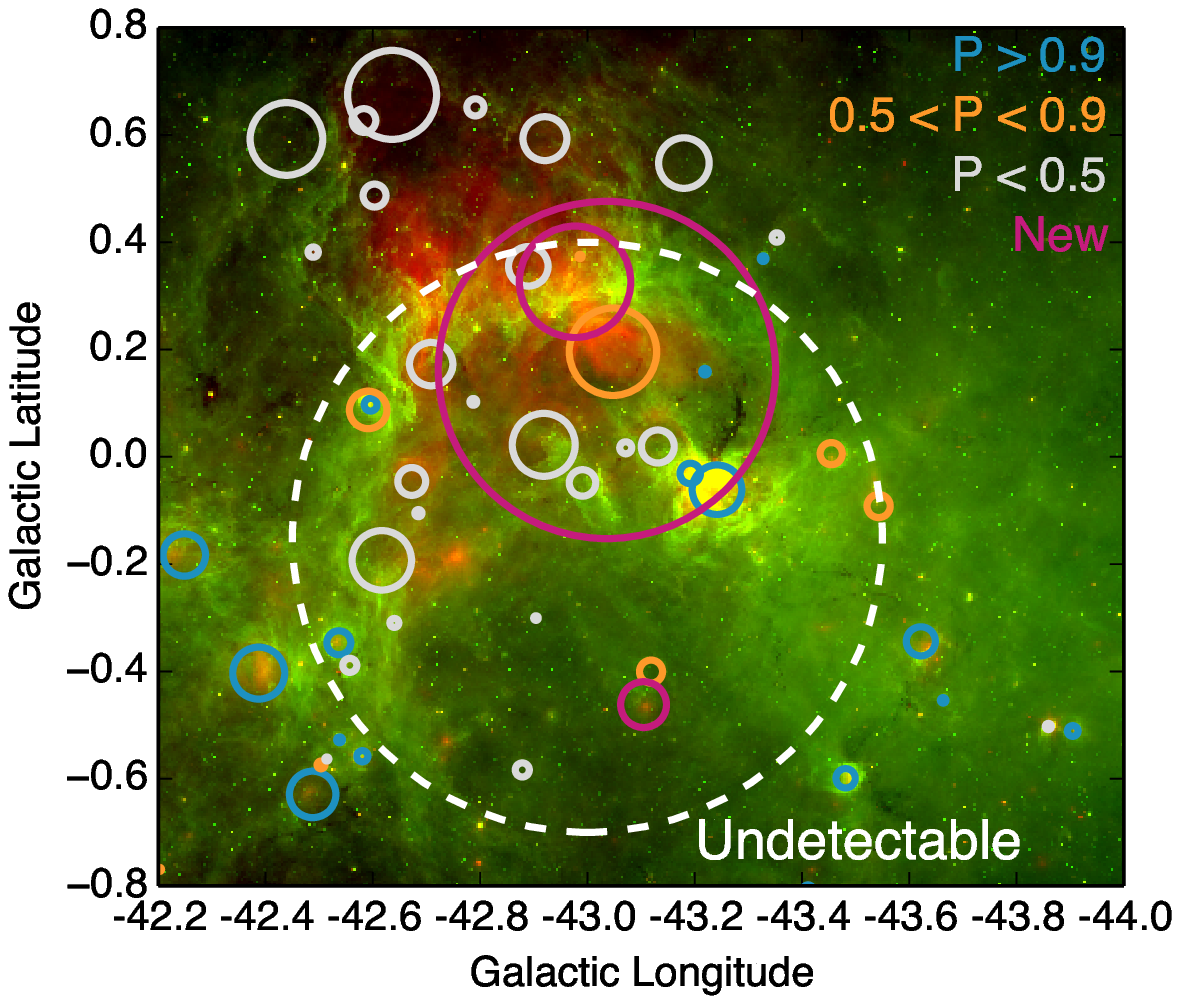}
\caption{Two fields with overabundances of low-probability bubbles in the MWP catalog.}
\label{fig:wide_fields}
\end{figure*}

To explore the differences between the high and low reliability portions of the MWP catalog, we have split the catalog into three roughly equally-sized groups: those where $P_{\rm bubble} < 0.5$, $0.5 < P_{\rm bubble} < 0.9$,
and $P_{\rm bubble} > 0.9$. Figure \ref{fig:dist_lon} shows the longitude distribution for each of these groups. The distributions are mostly similar, except at the three longitudes marked with vertical lines. These longitudes exhibit an excess of low-probability bubbles relative to the other groups.  All three fields are in fact sites of large, degree-scale giant \hii\, regions -- two are shown in Figure \ref{fig:wide_fields}. 

The excess of low-probability objects towards $\ell=-43^\circ$ coincides with a large bubble complex, containing bubbles S108-S111 in \cite{Churchwell06}. The region at $\ell=-55^\circ$ coincides with the Dragonfish nebula, studied by \cite{Rahman11}. Finally, the $\ell=-62^\circ$ region contains the G305 star forming complex studied by \cite{Hindson12}.

To determine whether the overabundance of low-probability sources near giant \hii\, regions indicates a failure of the probabilistic scoring or a bias in citizen scientist annotations, the lead author manually inspected the objects in the $\ell=-61^\circ$ bin. The sum of bubble probabilities for the 25 MWP objects in this bin is 8.09, and implies that an expert would identify $\sim 8$ of these objects as real bubbles if the probabilistic score is correct. Roughly 5-6 of these objects are compelling bubble candidates, which suggests that the excess of low-probability bubbles is not due to an overly conservative scoring strategy. Instead, this points to a real citizen scientist bias towards over-identification of bubbles in active regions like this. The central source in each of these regions creates a large, complex cavity in the ISM, and fluoresces much of the surrounding material as well. This leads to an abundance of coincidental, arc-like features misidentified as bubbles. The scores provided by Brut are able to identify this problem, and exclude the interlopers.

While citizen scientists are prone to over-identify bubbles towards giant \hii\, regions, they are unlikely to identify these large regions themselves. This might be because such objects are too irregular, or simply too big to see well in any single image served during the MWP. The widest zoom levels at which Brut scans through image data is wider than the widest images seen by citizen scientists, and it recovers some of these regions. For example, Figure \ref{fig:wide_fields} shows in magenta the new regions which Brut scores $>0.2$ on a blind search, that have no match in the MWP catalog. Each panel includes a new detection that is bigger than the biggest previously-cataloged objects. 

Still, Brut does not identify the large \hii\, regions themselves as bubbles. This is due to the fact that these large regions lie close to the $b = \pm 1^\circ$  boundary of {\sc glimpse} coverage, and Brut is unable to extract proper feature vectors at large scales (see Section \ref{sec:feature_edges}). For example, the dotted white circles in Figure \ref{fig:wide_fields} show plausible definitions for the large \hii\, regions. Both of these fields are undetectable, because they reside too close to the edge of latitude coverage. Thus, Brut converges on smaller detections. A feature extraction method that did not rely on pixels above and below a bubble might well be able to recover large bubbles close to latitude boundaries. 

In discussing the longitude distribution of bubbles, \cite{Simpson12} noted a dip in bubble counts at $|\ell| < 10^\circ$. They postulated that this might be because confusion in this busier region makes identification more difficult. However, this dip is most apparent for low-probability objects. In other words, the dip in the longitude distribution is not driven by a decrease in completeness, but rather by an increase in reliability. Perhaps, due to the complex background of emission towards the Galactic center, citizen scientists were less likely to notice subtle, coincidentally arc-like structures in the ISM.

\begin{figure}[h!]
\includegraphics{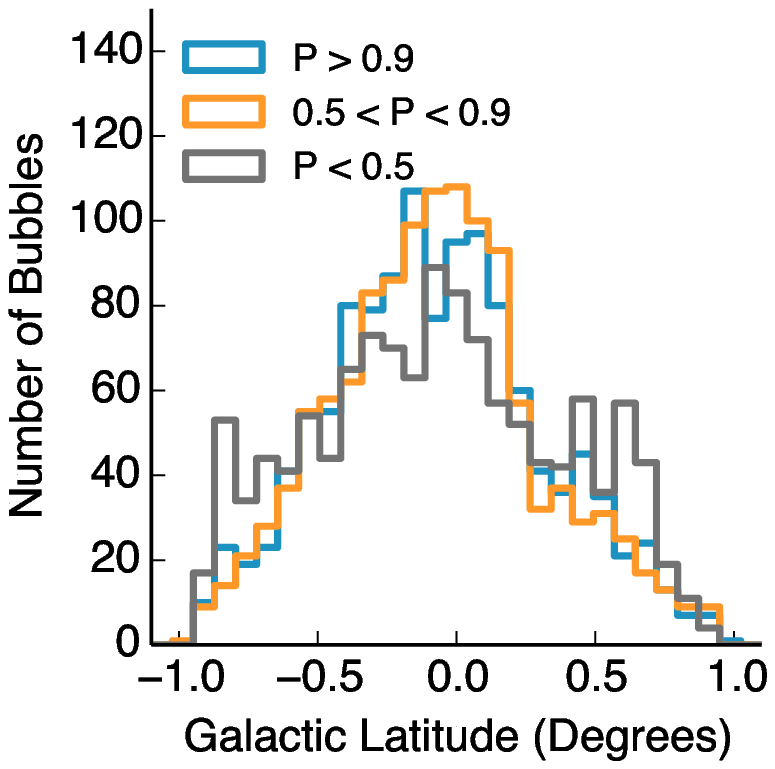}
\caption{The latitude distribution for the MWP catalog, partitioned according to the bubble probability score.
Low-probability objects have a slightly wider distribution.}
\label{fig:dist_lat}
\end{figure}

Figure \ref{fig:dist_lat} shows the latitude distribution for each subsample. The lowest-probability objects display a slightly broader distribution, particularly evident at $|b| > 0.5^\circ$. Because the ambient intensity field falls off quickly away from the mid-plane, we suspect random arc-like ISM structures are more likely to catch the eye in these regions. 

\begin{figure}[h!]
\includegraphics{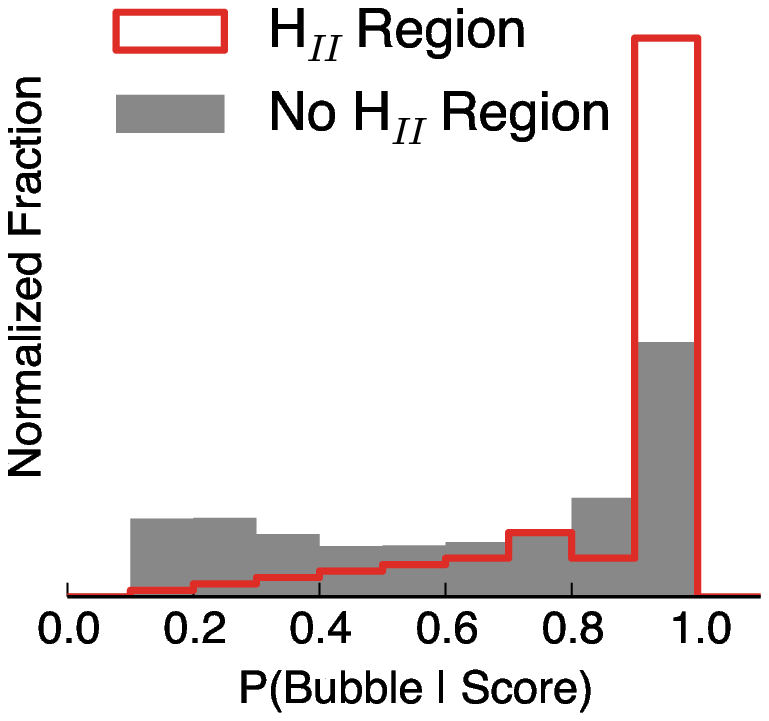}
\caption{The bubble probability distribution for sources with and without \hii\, region counterparts in the \cite{Anderson11} catalog.}
\label{fig:hii_score}
\end{figure}

Of the 3661 objects with bubble probability scores, 126 coincide spatially with \hii\, regions from \cite{Anderson11}. As Figure \ref{fig:hii_score} shows, objects associated with these \hii\, regions are strongly skewed towards higher bubble probabilities. This correlation strengthens the argument that the probability score successfully identifies dynamical objects.

\begin{figure}[h!]
\includegraphics[width=3in]{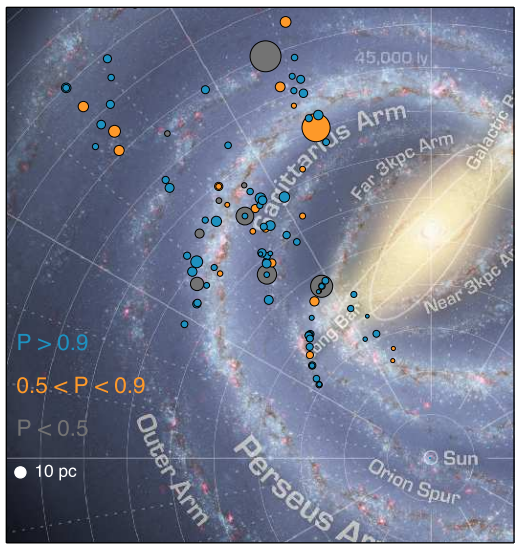}
\caption{Positions and sizes of bubbles with kinematic distance measurements from \cite{Anderson11}. The background image is an artist's schematic rendition of the Galaxy's spiral structure (credit NASA/JPL-Caltech/R. Hurt)}
\label{fig:dist_map}
\end{figure}

\cite{Anderson12} reports disambiguated kinematic distances for the \hii\, regions in their catalog. Figure \ref{fig:dist_map} shows the position and sizes of the bubbles associated with \hii\, regions with known distances, overlaid on a schematic map of the Galaxy's spiral structure. Four of the low-probability objects associated with \hii\, regions have large angular and linear diameters. However, from manual inspection none of these regions are compellingly associated with a real bubble in the image data. Thus, their match to \hii\, regions seems to be coincidental and arbitrary.

Uncertainties on these kinematic distances were calculated by \cite{Anderson12}, by considering uncertainties in the rotation curve model, large-scale non-circular motions, and standard parameters in the distance calculations, such as the Sun's distance to the Galactic Centre and the solar orbital speed. The mean uncertainty on the full HRDS sample is 0.5 kpc; they find that 90\% of loci in their sample have $\leq 20\%$ combined uncertainty from these three effects. Furthermore, they do not find the uncertainties to have a significant effect on the galactic distribution of their sample. 

While bubbles cluster around the Sagittarius arm, there are two clusters of bubbles in inter-arm regions -- one beyond the outer arm, and another apparent stream of bubbles around 5kpc, from $30^\circ < \ell < 60^\circ$. A bubble is a broad observational definition that covers several astrophysical phenomena, including \hii\, regions, non-ionized cavities from B star winds, and supernova remnants. Bubbles located on or between spiral arms may well correspond to different astrophysical classes or evolutionary stages. Furthermore, the detectability of bubbles is a complicated mixture of an object's size and brightness as a function of time, as well as the amount of unrelated structure along the line of sight to each object. Population synthesis studies which simulate realistic galactic environments and predict distributions akin to Figure \ref{fig:dist_map} might be able to constrain the relative mixture -- and dynamical importance -- of different classes of bubbles \citep{Robin03,Robitaille12}.

\subsection{Evidence for Triggering}

Bubbles are frequently studied in the context of triggered star formation. The material excavated by bubble-blowing stars might induce subsequent generations of star formation. Two main mechanisms for triggered star formation have been proposed: in the collect and collapse model \citep{Whitworth94, Dale07}, material gathered along bubble rims eventually passes a critical density and undergoes gravitational fragmentation and collapse. In the radiatively-driven implosion model, the wind or ionization shock front collides with pre-existing but stable cores, and the resulting compression triggers collapse \citep{Bertoldi89}.

In any particular region, finding clear evidence for triggered star formation is difficult. The typical approach is to identify an overdensity of young stars within or on a bubble rim, and/or to look for an inverse age gradient with bubble radius \citep{Deharveng05, Zavagno06, Koenig08}. Such an analysis is often confounded by small numbers of YSOs and ambiguous line-of-sight distances and stellar ages. Furthermore, it is often unclear whether spatial correlations between bubbles and YSOs imply a triggered relationship between the two, since star formation is a naturally clustered process \citep{Lada03}.

Such problems can partially be addressed by correlating YSOs with large bubble catalogs like the MWP catalog -- this does not disambiguate correlation from causation, but it can overcome problems related to small-number statistics. \cite{Thompson12} first applied such an analysis using the \cite{Churchwell06} catalog, and \cite{Kendrew12} later repeated this on the MWP catalog. This analysis computes an angular correlation function \citep{Landy93, Bradshaw11}, defined by

\begin{equation}
w(\theta) = \frac{N_{BY} - N_{BR_Y} - N_{R_BY} + N_{R_BR_Y}}{N_{R_BR_Y}}
\label{eq:corr}
\end{equation}
where $N_{\alpha \beta}$ represents the number of pairs of objects from catalogs $\alpha$ and $\beta$ with a separation of $\theta$, $B$ is a bubble catalog, $Y$ is a YSO catalog, and $R_B$ and $R_Y$ are randomly-distributed bubble and YSO catalogs. These random locations are chosen to preserve the approximate latitude and longitude distribution of each class of objects, but are otherwise uniformly distributed. Informally, $w(\theta)$ represents the excess likelihood of finding a YSO at a particular distance $\theta$ from a bubble, relative to what is expected from a random distribution of bubbles and YSOs. We normalize the angular offset $\theta$ by the radius of each bubble, such that $w(\theta)$ traces the excess YSOs as a function of offset in bubble radii. An ideal signature of triggered star formation, then, would be a local maximum of $w(\theta)$ at $\theta = 1$. 

\begin{figure}[h!]
\includegraphics[width=3.1in]{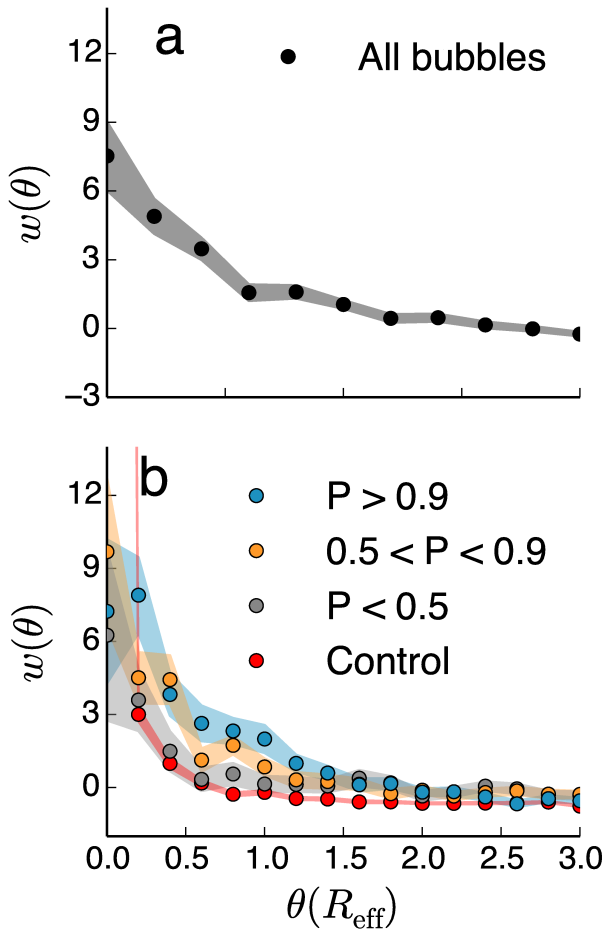}
\caption{Angular correlation function (Equation \ref{eq:corr}) between bubbles and sources in the RMS catalog of YSOs. Error estimates (shading) are derived via bootstrap resampling from the bubble catalog. Top: the correlation function for all bubbles. Bottom: the correlation function for the
low (black), medium (orange), and high-reliability bubbles (blue). The red control curve is obtained by repositioning each bubble to coincide with a random YSO -- it measures the general clustering of YSOs.}
\label{fig:trigger}
\end{figure}

We reproduce the Figure 15 of \cite{Kendrew12} in Figure\ref{fig:trigger}a. This shows the angular correlation function between the MWP Large catalog and RMS catalog of YSOs and compact \hii\, regions \citep{RMS}, as a function of normalized bubble radius. The main signal is a decaying correlation, indicative of the fact that star formation occurs in clustered environments. The prediction from triggered star formation is that there should be an additional peak at $\theta \sim 1$ bubble radius. Such a signal is not obvious in this figure, though \cite{Kendrew12} report evidence for a peak for the largest bubbles in the MWP catalog. Likewise, \cite{Thompson12} report a clearer peak when considering the expert-selected bubbles in \cite{Churchwell06}. 

In Section \ref{sec:expert}, we demonstrated that roughly 30\% of objects in the MWP catalog are interlopers -- random ISM structures incorrectly tagged as bubbles. Furthermore, the relative false positive rate increases towards giant \hii\, regions -- precisely the regions where one might expect triggered star formation to be most apparent. Interlopers might significantly dilute bubble/YSO correlations in the MWP catalog. Fortunately, our bubble probability scores allow us to identify many of these interlopers, yielding a higher reliability catalog that still has 3x as many bubbles as used by \cite{Thompson12}.

Figure \ref{fig:trigger}b shows the same correlation function, for the subsamples partitioned by bubble probability. This reveals an additional excess from $0.5 < \theta < 1$ bubble radius, whose strength increases with bubble probability. This is similar to the trend with bubble size that \cite{Kendrew12} reported, but the signal here is both stronger and due to a different subsample of bubbles -- the size distribution of bubbles does not vary significantly between the three probability bins.

Star formation is a naturally clustered process. The curves in Figure \ref{fig:trigger} trace the excess of RMS YSOs near bubbles relative to a purely random distribution of objects -- they do not measure the excess density of YSOs relative to other star formation regions. The red ``Control'' curve in Figure \ref{fig:trigger} addresses this. This curve was obtained by repositioning each bubble to the location of a random RMS YSO, and re-running the analysis. In other words, the curve shows the natural clustering of YSOs relative to each other. The left-most point of the curve is very large, which is an artifact of the fact that each bubble lines up exactly with the YSO it was repositioned to -- this creates a strong overdensity at very small $\theta$. Beyond this point, however, the curve resembles the $P < 0.5$ subset. The fact that the $P > 0.9$ bubbles fall significantly above the control curve indicates that YSOs are clustered around high-probability bubbles even more strongly than they cluster around one another on average. However, this clustering analysis cannot determine whether the overdensity around $P > 0.9$ bubbles is the result of bubble-triggered star formation, or rather an indication that bubbles form in particularly active star forming regions. 

\section{Blind Search}
\label{sec:blind}
The previous section focused on using Brut to reassess bubbles previously identified by citizen scientists. We've demonstrated that Brut is successful at identifying the high-reliability subset of the MWP catalog and, conversely, at flagging probable interlopers in the catalog. The result is a purer statistical sample of bubbles in the Milky Way.

Now we consider the possibility of finding bubbles missing from the MWP catalog. By scanning through all {\sc glimpse}/{\sc mipsgal} data, Brut can be used to conduct a blind search for new bubbles. Discovering bubbles without knowing the citizen-science hit-rate at each location is a harder task; Brut does not benefit from complementary information about how citizen scientists classify a particular region. However, this task is relevant to future projects where machine learning techniques assist manual search. For applications where exhaustive human search is infeasible, machine learning algorithms can conduct exhaustive searches and flag interesting candidate objects for human attention or followup observation.

We performed a blind search with Brut as follows: starting at a minimum field of view of 80'', we scanned the entire {\sc glimpse}/{\sc mipsgal} survey area of $|\ell| < 65^\circ$, $|b| < 1^\circ$. Each field was offset from its neighbor by 20\% of the field size. At each location, we computed the Brut score. After scanning the entire survey area, we increased the field of view by 25\%, and re-scanned the Galactic plane at this larger scale. This process was repeated up to a maximum field of view of 1 degree. In total, this produced approximately 35 million classifications. Of these, we extracted 58,294 fields with Brut scores greater than 0.2, and merged these following the procedure in Section \ref{sec:build}. 

\begin{figure}[h!]
\includegraphics{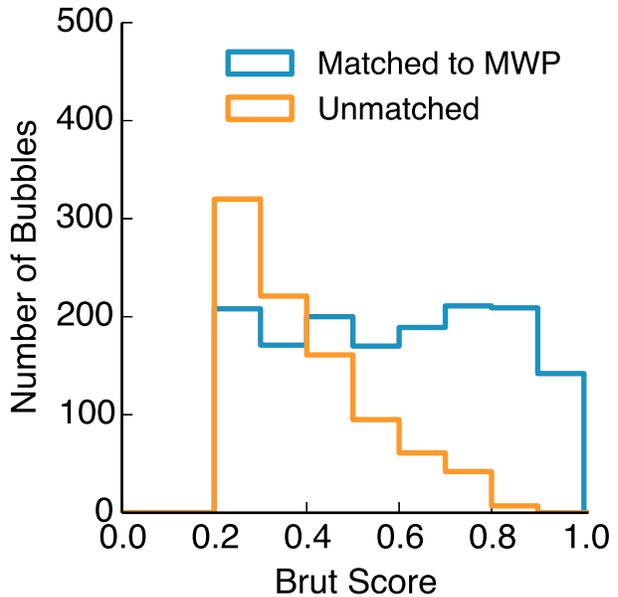}
\caption{The distribution of Brut Scores for bubble candidates identified during the blind search.}
\label{fig:new_score}
\end{figure}

This process yielded a list of 2407 distinct bubble candidates. According to Figure \ref{fig:expert_uniform_score}, an Expert is about 50\% likely to judge a region with a Brut score of 0.2 as a bubble. Thus, this candidate sample is very generous, and probably includes a substantial interloper population. Using the fit to Figure \ref{fig:expert_uniform_score}, the summed probability for all objects -- and hence the expected number of genuine bubbles in this sample -- is 1243. 1500 objects in the blind search have counterparts in the MWP catalog, and 907 do not. Figure \ref{fig:new_score} shoes the Brut score distribution for objects with and without MWP counterparts. Note that objects with no MWP counterpart are skewed towards lower scores, and the majority of these are interlopers. Brut's blind search does not reveal any significant statistical incompleteness in the MWP catalog.

\begin{figure*}
\includegraphics{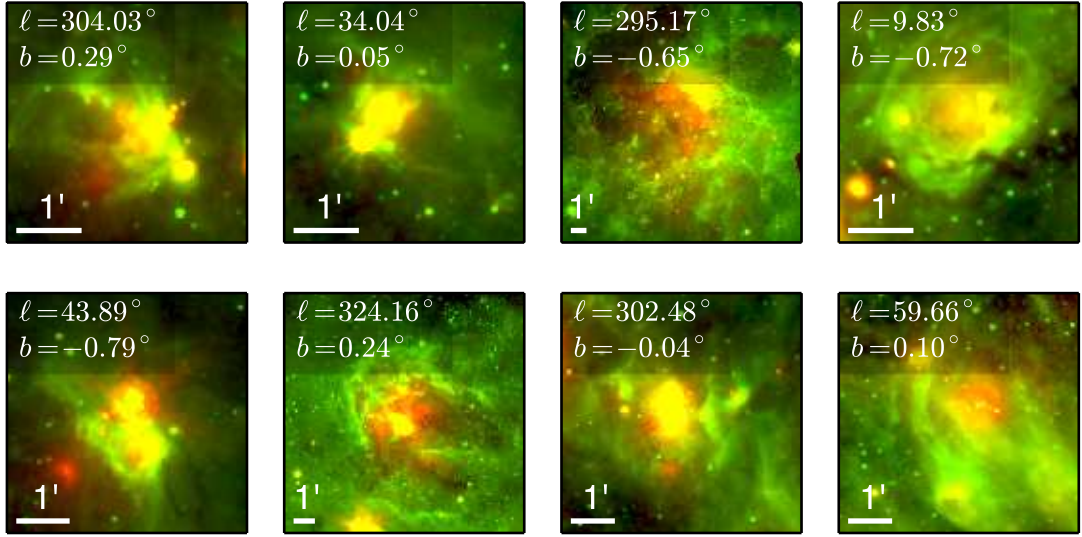}
\caption{Eight bubbles not present in the Milky Way Project catalog, discovered by Brut during a blind search.}
\label{fig:new_gallery}
\end{figure*}

Still, Brut does recover a handful of genuine bubbles missing from the MWP catalog. The easiest way to find these is to sort the 907 unmatched bubble candidates by Brut score, and manually examine the highest-scoring regions. Figure \ref{fig:new_gallery} presents 8 of the most compelling bubble candidates with no MWP counterparts -- these bubbles are among the 70 highest-scoring regions with no MWP match. We have examined the original MWP images associated with each region and find that, in most cases, these bubbles are sufficiently close to a bright source that they are hard to see. Because Brut locally adjusts the contrast of each field when building feature vectors, it overcomes this difficulty. At the same time, these eight objects represent $\sim 10\%$ of the high-scoring candidates we examined. The remaining objects are false positives -- many of them smaller substructures of larger bubbles in the MWP catalog, or ambiguous sources of 24\um\, nebulosity. Brut is not discriminating enough to find bubbles missed by the MWP on its own. However, it \textit{is} effective at generating promising candidates to followup on.

We can't rule out the possibility of bubbles missed by both Brut and the MWP -- it is possible, for example, that Brut could have learned a selection bias present in the training data. However, we conclude from this exercise that Brut's ability to identify bubble candidates in a blind search is comparable to citizen scientists, and such techniques can be useful as a way to pre-process large datasets. Furthermore, we note that the combined efforts of Brut and the MWP yield a much larger catalog of high-reliability bubbles than the previous catalogs curated by professional astronomers \citep{Churchwell06, Churchwell07}. 

\section{Next Steps}
\label{sec:next_steps}
The success of Brut demonstrates the potential synergies that exist between machine learning, professional scientists, and citizen scientists. Note the complementary strengths and weaknesses of each resource:

\begin{enumerate}
\item Professional scientists are best-suited to perform nuanced classification tasks that require domain-specific knowledge. They are also the most resource-limited. 
\item Citizen scientists outnumber professional scientists by orders of magnitude (in the case of Bubble detection, the factor is nearly 10,000:1). They are equally capable with the generic aspects of pattern recognition, but do not possess the domain expertise of professionals. Furthermore, curious citizen scientists are well-situated for serendipitous discovery of unusual objects \citep{Lintott09, Cardamone09}.
\item Supervised machine learning algorithms have no a-priori pattern recognition ability, and require external training. However, once supplied with this information, computer-driven analyses are reproducible and extremely scalable.
\end{enumerate}

Brut leverages the information provided by astronomers and citizen scientists alike. Citizen scientist input was used for training, and a smaller dataset of high-quality expert information set was used to convert Brut's classifications into calibrated bubble probabilities. The result is a classifier that is both more precise than the raw citizen scientist catalog, and more complete than the best expert-assembled catalog.

Searching \emph{Spitzer} images for bubbles is a small enough task that citizen scientists were able to perform an exhaustive search. Consequently, the MWP catalog appears to contain most of the bubbles that one can hope to identify from these images. Brut's main benefit is in providing an independent, probabilistic assessment of the MWP catalog, identifying interlopers in the catalog, and adding a small number of bubbles missed by citizen scientists -- particularly bubbles near bright objects.

However, one can hypothetically envision tools like Brut assisting professional and citizen scientists in real time. For searches for rarer objects in larger datasets, exhaustive human search is infeasible -- both due to boredom and prohibitive data sizes. Had Brut been trained at the start of the MWP, it would quickly have been able to eliminate many regions as devoid of bubbles. Citizen scientists could have spent more time classifying more ambiguous regions, which is where their input is most valuable (and where the task is most interesting). These ideas are explored in more depth by \cite{MSR}, and will become increasingly important as data continues to grow. 

Similarly, Brut could easily incorporate additional sources of information. For example, far-infrared observations from \emph{Herschel} constrain the column density and temperature of dust in the vicinity of bubbles \citep{Anderson12b}. This additional information can further disambiguate real bubbles from other cavities in the ISM. A followup investigation could supplement Brut's feature vectors with features extracted from \emph{Herschel} data, and retrain the classifiers. This is a promising approach to search for bubbles in the outer galaxy, since \emph{Spitzer} did not systematically survey the outer Galaxy at 8 or 24 \um\, and citizen scientists have not (yet) surveyed this region.

\section{Conclusion}
\label{sec:conclusion}
We have developed an automatic bubble detector, Brut, using the Random Forest classification algorithm in tandem with a catalog of citizen scientist-identified bubbles in our galaxy. This algorithm is effective at detecting bubbles in \emph{Spitzer} images. By comparing the confidence scores that Brut computes with expert classifications of a small set of images, we are able to estimate the probability that any given region in \emph{Spitzer} {\sc glimpse} and {\sc mipsgal} data contains a bubble. We have used Brut to re-assess the objects in the MWP catalog, and also to perform a full search over {\sc glimpse} and {\sc mipsgal} images for new bubbles.  Several insights have emerged from this analysis:
\begin{enumerate}
\item Roughly 30\% of the objects in the MWP catalog are interlopers -- structures which a majority of experts would not consider to be likely \hii\, regions or wind blown bubbles.
\item Brut is able to identify which objects are probable interlopers, and likewise to identify highly probable bubbles. Compared to the MWP catalog as a whole, high-probability bubbles have a narrower latitude distribution, and are nearly 5 times more likely to be associated with \hii\, regions identified by \cite{Anderson11}.
\item The MWP catalog has a higher concentration of low-probability bubbles near giant \hii\, regions, which fluoresce the surrounding ISM and reveal more coincidental circular structures. Citizen scientists are prone to identify these regions as bubbles, whereas experts and Brut do not.
\item High probability bubbles  exhibit stronger excesses of YSOs and compact \hii\, regions along and interior to bubble rims -- a prediction of triggered star formation theories.
\end{enumerate}

Image classification remains a difficult problem in many contexts, and techniques like Brut are not yet as good as expert human analysis. However, Brut demonstrates that automated techniques are a valuable complement to manual search.  Combining human and machine searches is the most promising avenue for scaling tasks like astrophysical image classification to very large datasets.

This research made use of the Scikit-Learn \citep{Pedregosa11}, Astropy \citep{Astropy}, and WiseRF software packages. WsieRF is a product of wise.io, Inc., and we thank them for making an academic license freely available. JPW acknowledges support from NSF-AST1108907. This work is based on observations made with the Spitzer Space Telescope, which is operated by the Jet Propulsion Laboratory, California Institute of Technology under a contract with NASA.

\appendix{
\section{Expert Survey}
Figure \ref{fig:expert_form} shows the interface experts used to provide classifications for the expert survey. Each object is shown 6 times, at 3 zoom levels (columns) and two contrast settings (rows). Experts were asked to click one of the three buttons to classify the object as a non-bubble, ambiguous/irregular bubble, or clear bubble. To minimize the effects of fatigue, each user classified objects in a different, random order.

Figure \ref{fig:mwp_survey_gallery} shows the objects in the survey that were drawn randomly from the MWP catalog. Each image corresponds to the upper-left image of the survey form. Similarly, Figure \ref{fig:uniform_survey_gallery} shows the objects in the uniform sample. Recall that these were selected to be uniformly distributed in Brut's confidence score.

\begin{figure}
\includegraphics[width=5in]{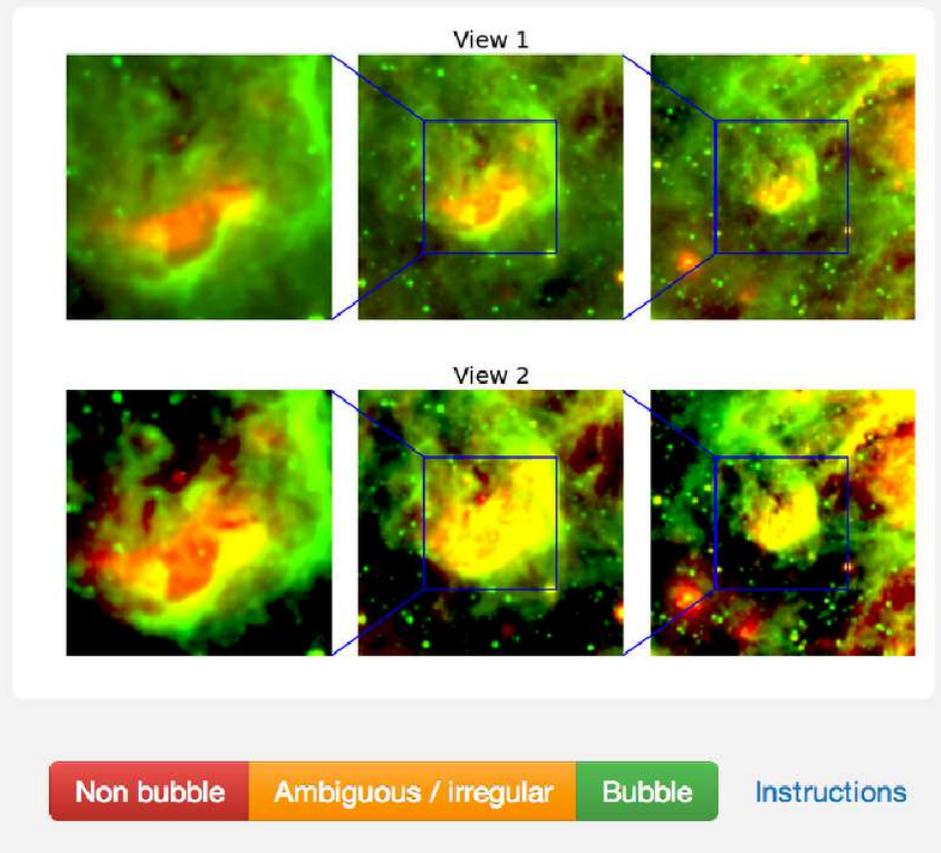}
\caption{A page from the expert survey, showing a possible bubble at 3 zoom levels and 2 contrast settings.}
\label{fig:expert_form}
\end{figure}

\begin{figure}
\includegraphics[width=7in]{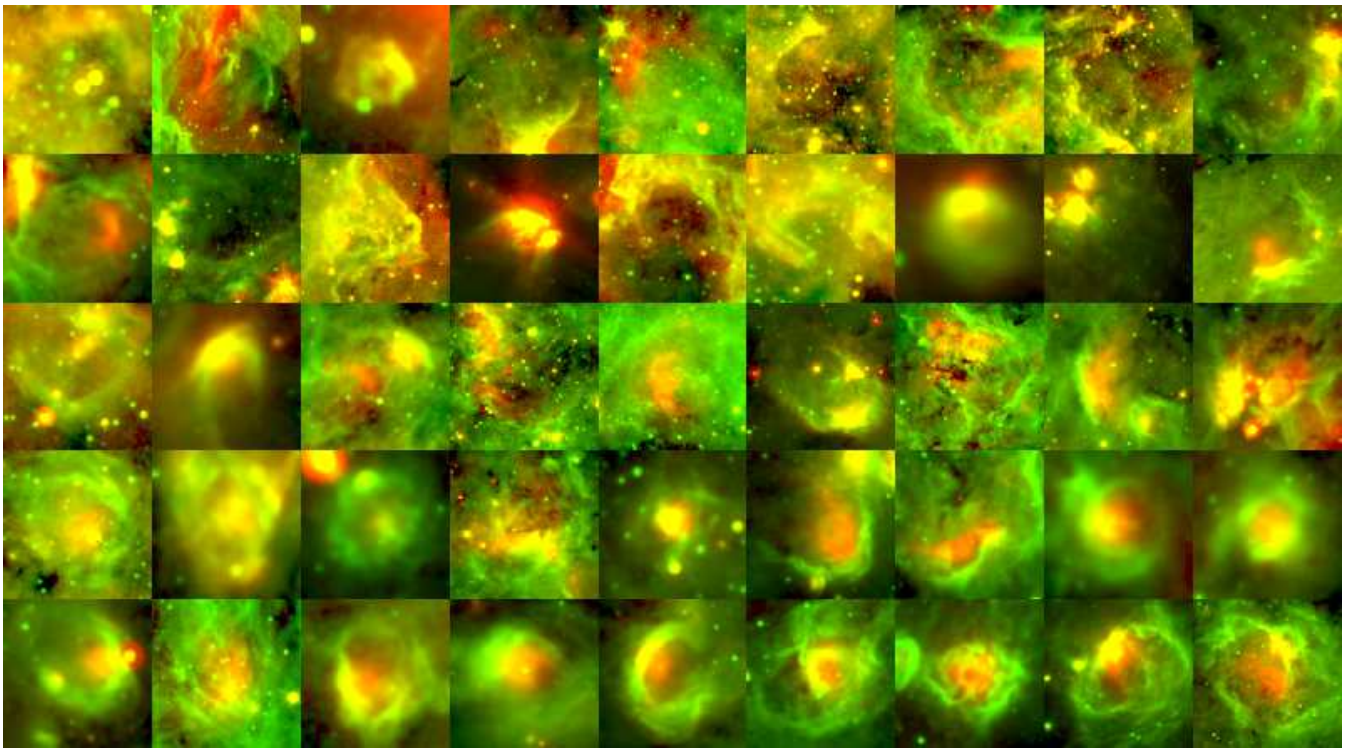}
\caption{The objects in the expert survey selected from the MWP catalog. Objects are sorted by Brut's confidence score.}
\label{fig:mwp_survey_gallery}
\end{figure}

\begin{figure}
\includegraphics[width=7in]{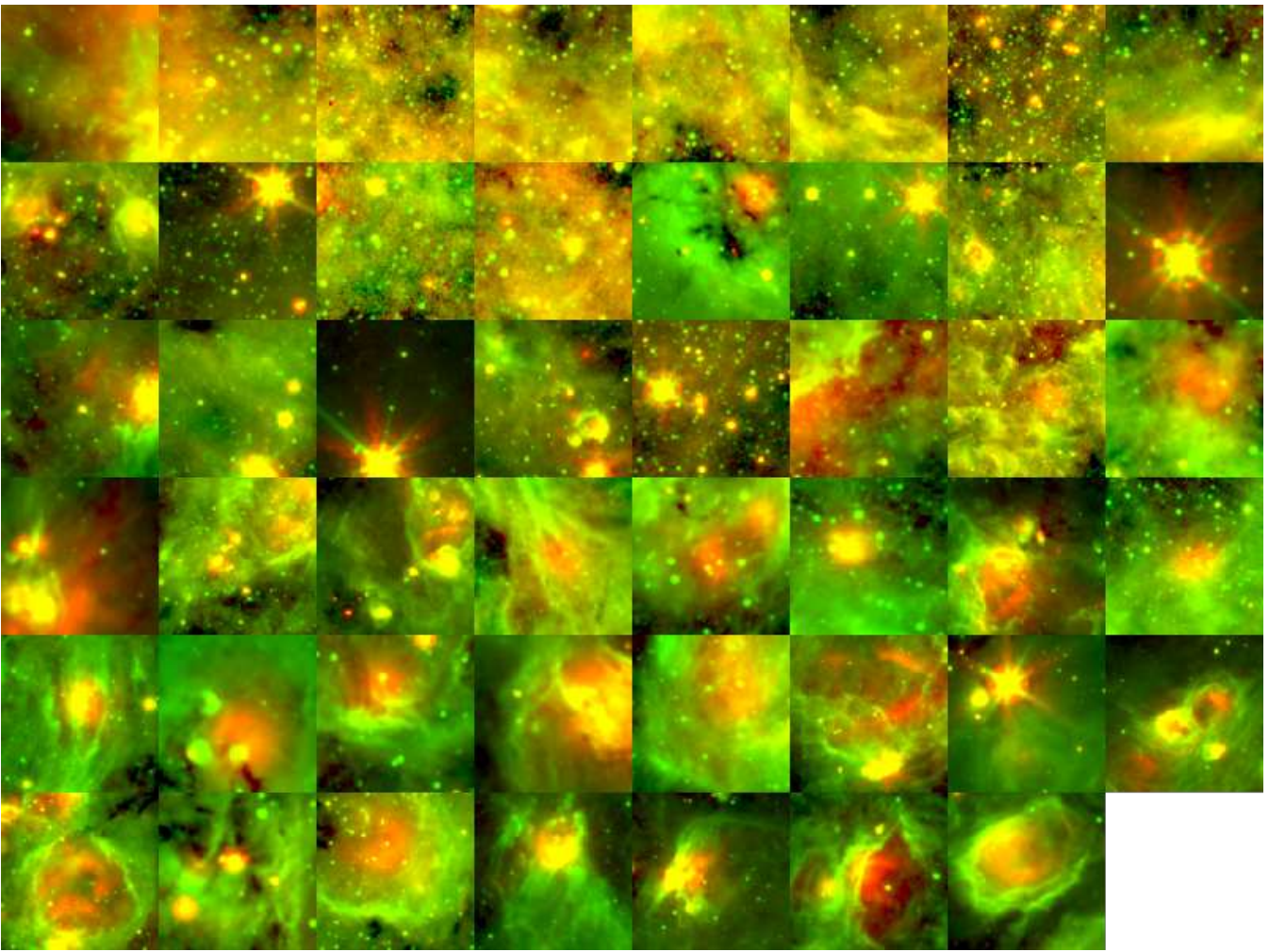}
\caption{The same as Figure \ref{fig:mwp_survey_gallery}, but for objects from the uniform sample.}
\label{fig:uniform_survey_gallery}
\end{figure}

}
\input{prob_table_head.tex}

\end{document}

%% file: prob_table_head.tex
\begin{deluxetable}{rrrrrrr}
\tablecaption{Bubble probabilities for objects in the Milky Way Project catalog. Non-finite probabilities correspond to objects near the edge of the data, which Brut cannot analyze.\label{table:prob_table}}
\tablehead{\colhead{$\ell$} & \colhead{$b$} & \colhead{PA} & \colhead{Semimajor axis} & \colhead{Semiminor axis} & \colhead{Hit Rate} & \colhead{P(Bubble)}\\ \colhead{deg} & \colhead{deg} & \colhead{deg} & \colhead{arcmin} & \colhead{arcmin} & \colhead{ } & \colhead{ }}
\startdata
312.9748272 & -0.4354279 & 23 & 1.7 & 1.7 & 0.53 & 1.000 \\
309.059036 & 0.1660606 & 23 & 3.7 & 3.5 & 0.437 & 0.999 \\
354.182135 & -0.0527919 & 10 & 2.0 & 1.8 & 0.524 & 1.000 \\
348.2597731 & 0.477404 & 25 & 3.9 & 3.8 & 0.459 & 0.999 \\
345.4839785 & 0.4011382 & 17 & 1.7 & 1.7 & 0.536 & 1.000 \\
13.7330073 & -0.0159981 & 18 & 3.4 & 3.3 & 0.543 & 0.999 \\
5.8986592 & -0.4548185 & 25 & 4.0 & 3.9 & 0.38 & 0.993 \\
18.192723 & -0.4001685 & 19 & 1.9 & 1.7 & 0.535 & 0.999 \\
18.2606111 & -0.2967184 & 17 & 1.7 & 1.6 & 0.672 & 1.000 \\
331.0108867 & -0.1484733 & 19 & 2.2 & 2.1 & 0.502 & 0.999 \\
\enddata
\end{deluxetable}